\documentclass[11pt]{article}

\usepackage{epsfig,subfigure}
\usepackage{amssymb, amsmath}
\usepackage{color}
\usepackage{hyperref}
\usepackage{graphicx}
\usepackage{color}

\usepackage{amssymb}
\usepackage{amsmath,amsfonts,amssymb}

\usepackage{verbatim}
\usepackage{stfloats}
\usepackage[bookmarks=false]{}

\newtheorem{theorem}{Theorem}

\newtheorem{lemma}{Lemma}

\begin{document}
\date{}
\title{Aligned Image Sets under Channel Uncertainty: Settling  a Conjecture  by Lapidoth, Shamai and Wigger on the Collapse of Degrees of Freedom under Finite Precision CSIT}
\author{ \normalsize Arash Gholami Davoodi and Syed A. Jafar \\
{\small Center for Pervasive Communications and Computing (CPCC)}\\
{\small University of California Irvine, Irvine, CA 92697}\\
{\small \it Email: \{gholamid, syed\}@uci.edu}
}
\maketitle

\begin{abstract}
A  conjecture made by Lapidoth, Shamai and Wigger at Allerton 2005  (also an open problem presented at ITA 2006) states that the degrees of freedom (DoF) of a two user  broadcast channel, where the transmitter is equipped with $2$ antennas and each user is equipped with $1$ antenna, must collapse under finite precision channel state information at the transmitter (CSIT). That this conjecture, which predates interference alignment, has remained unresolved, is emblematic of a pervasive lack of understanding of the degrees of freedom of wireless networks---including interference and $X$ networks---under channel uncertainty at the transmitter(s). In this work we prove that the conjecture is true in all non-degenerate settings (e.g., where the probability density function of unknown channel coefficients exists and is bounded). The DoF collapse even when  perfect channel knowledge for one user is available to the transmitter. This also settles a related recent conjecture by Tandon et al. The key to our proof is a bound on the number of codewords that can cast  the same image (within noise distortion) at the undesired receiver whose channel is subject to finite precision CSIT, while remaining resolvable at the desired receiver whose channel is precisely known by the transmitter. We are also able to generalize the result along two directions. First, if the peak of the probability density function is allowed to scale as $O((\sqrt{P})^\alpha)$, representing the concentration of probability density (improving CSIT) due to, e.g.,  quantized feedback at rate $\frac{\alpha}{2}\log(P)$, then the DoF are bounded above by $1+\alpha$, which is also achievable under quantized feedback. Second, we generalize the result to the $K$ user broadcast channel with $K$ antennas at the transmitter and a single antenna at each receiver. Here also the DoF collapse under non-degenerate channel uncertainty. The result directly implies a collapse of DoF to unity under non-degenerate channel uncertainty  for the general $K$-user interference and $M\times N$ user $X$ networks as well, for which the best known outer bounds  under non-degenerate channel uncertainty (except for essentially degraded settings) prior to this work were $\frac{K}{2}$ and $\frac{MN}{M+N-1}$ (same as with perfect CSIT).
\end{abstract}
\newpage

\section{Introduction}
Interference alignment studies \cite{Jafar_FnT} have spurred  much interest in the degrees of freedom (DoF) of wireless communication networks. While much progress has been made under the assumption of perfect channel knowledge, the degrees of freedom under channel uncertainty at the transmitters have remained mostly a mystery. A prime example is the, heretofore unresolved, conjecture by Lapidoth, Shamai and Wigger from the Allerton conference in 2005  \cite{Lapidoth_Shamai_Wigger_BC}, also  featured at the ``\emph{Open Problems Session}" at the Inaugural Information Theory  and its Applications (ITA) workshop in 2006 \cite{Lapidoth_Shamai_Wigger_ITA}, which claims that the DoF  collapse under finite precision channel state information at the transmitter (CSIT). Specifically, Lapidoth et al. conjecture that the DoF of a 2 user multiple input single output (MISO) broadcast channel (BC) with 2 antennas at the transmitter and 1 antenna at each of the receivers, must collapse to unity (same as single user) if the probability distribution of the channel realizations, from the transmitter's perspective, is sufficiently well behaved that the differential entropy rate is bounded away from $-\infty$. The condition excludes not only settings where  some or all channel coefficients are perfectly known,  but also scenarios where some channel coefficients are functions of  others, even if their values remain unknown.  The best DoF outer bound under such channel uncertainty, also obtained by Lapidoth  et al., is $\frac{4}{3}$. Deepening the mystery is the body of evidence on both sides of the conjecture. On the one hand, supporting evidence in favor of the collapse of DoF is available if the channel is essentially degraded, i.e., the users' channel vector directions are statistically indistinguishable  from the transmitters' perspective \cite{Huang_Jafar_Shamai_Vishwanath, Varanasi_noCSIT}. On the other hand, the idea of blind interference alignment  introduced by Jafar in  \cite{Jafar_corr}  shows that the  2 user MISO BC achieves $\frac{4}{3}$ DoF (which is also an outer bound, thus optimal), even without knowledge of channel realizations at the transmitter, provided that one user experiences time-selective fading and the other user experiences frequency-selective fading. Since the time-selective  channel is assumed constant across frequency and the frequency-selective channel is assumed constant across time, it makes some channel coefficients  functions of others (they are equal if they belong to the same coherence time/bandwidth interval), so that the model does not contradict the conjecture of Lapidoth et al. Thus, quite remarkably, this conjecture of Lapidoth, Shamai and Wigger, which predates interference alignment in wireless networks, has remained unresolved for nearly a decade.

Following in the footsteps of Lapidoth et al., subsequent works have made similar, sometimes even stronger conjectures, as well as partial attempts at proofs. For instance, the collapse of DoF of the MISO BC was also conjectured by Weingarten, Shamai and Kramer in \cite{Weingarten_Shamai_Kramer} under the finite state compound setting. However, this conjecture turned out to be too strong and was shown to be false by Gou, Jafar and Wang in \cite{Gou_Jafar_Wang}, and by Maddah-Ali in \cite{Maddah_Compound}, who showed that, once again, $\frac{4}{3}$ DoF are  achievable (and optimal) for almost all realizations of the finite state compound MISO BC, regardless of how large (but finite) the number of states might be. Since the differential entropy of the channel process is not defined (approaches $-\infty$) for the finite state compound setting, this result also does not contradict the conjecture of Lapidoth et al. A related refinement of the conjecture, informally noted on several occasions  (including by Shlomo Shamai at the ITA 2006 presentation) and mentioned most recently (although  in the context of i.i.d. fading channels) by Tandon, Jafar, Shamai and Poor  in \cite{Tandon_Jafar_Shamai_Poor} --- is that the DoF should collapse even in the ``PN" setting, where  perfect (P) CSIT is available for one of the two users, while no (N) CSIT is available for the other user. A valiant attempt at proving this conjecture is made in \cite{Hao_Rasouli_Clerckx}, but it turns out to be  unsuccessful because it relies critically on an incorrect use of the extremal inequality of \cite{Liu_Viswanath} under channel uncertainty.\footnote{Similar problems arise in \cite{Rassouli_Hao_Clerckx, Hao_Clerckx}. A simple counter example is the  MISO BC with finitely many channel states, where the same  arguments as used in these works  would imply a collapse of DoF to 1, whereas this setting is known to have $\frac{4}{3}$ DoF as shown in \cite{Gou_Jafar_Wang, Maddah_Compound}.} Thus the ``PN" conjecture  has also thus far remained  unresolved.

That these  conjectures remain unresolved, is emblematic of a broader lack of understanding of the DoF of wireless networks under non-degenerate forms of channel uncertainty. For instance, by extension, under non-degenerate channel uncertainty we also do not know the DoF of the vector broadcast channel with more than 2 users, or the DoF of interference networks, $X$ networks, cellular, multi hop, or two-way relay networks, with or without multiple antennas,   or any of a variety of settings with partial uncertainty, such as mixed \cite{Gou_Jafar, Sheng_Kobayashi_Gesbert_Yi} or alternating \cite{Tandon_Jafar_Shamai_Poor} channel uncertainty. Thus, the resolution of these conjectures is likely to have a broad impact on our understanding of the ``robustness" of the DoF of wireless networks. This is the motivation for our work in this paper.

\subsection{Overview of Contribution}
The main contribution of this work is to prove the conjecture of Lapidoth, Shamai and Wigger,  thereby closing the ITA 2006 open problem, as well as the ``PN" conjecture of Tandon et al., for all non-degenerate forms of finite precision CSIT, which includes all settings where density functions of the unknown channel realizations exist and are bounded. For all such settings, we show that the DoF collapse to unity as conjectured. Remarkably, this is the first result to show the total collapse of DoF under channel uncertainty without making assumptions of degradedness, or the (essentially) statistical equivalence of users.

Our approach is based on  bounding the expected number of codewords that are resolvable at their desired receiver whose images align (within bounded noise distortion) at the undesired receiver under finite precision CSIT. We show that this quantity is $\approx O((\log(P))^{n})$ where $n$ is the length of codewords, and $P$ is the power constraint which defines the DoF limit as $P\rightarrow \infty$. This is negligible relative to the total number of resolvable codewords, which is $\approx O(P^{nd/2})$ when the desired information is sent at rate $\frac{d}{2}\log(P)$, i.e., with DoF $d>0$  (normalization by $\frac{1}{2}\log(P)$ is because we deal with real channels). The difference between the entropy contributed by any set of codewords at their desired receiver (desired DoF) and the entropy contributed by the same set of codewords at the undesired receiver (DoF consumed by interference) tends to zero in the DoF sense. Under non-degenerate channel uncertainty, it is not possible to utilize the DoF at the desired receiver without sacrificing the same number of DoF at the undesired receiver due to interference. Therefore, the DoF are bounded above by unity, the same as with a single user.

We  also generalize this result in two directions. First, we extend it to include CSIT that improves as $P\rightarrow\infty$, e.g., through quantized feedback at rate $\frac{\alpha}{2}\log(P)$, so that the probability density function of unknown channel coefficients concentrates around the correct realizations. This refinement of CSIT is captured by the growth in the peak value of the probability density function. We show that if the peak of the probability density function of unknown channel coefficients grows no faster than $O(P^\frac{\alpha}{2})$, representing e.g., improving channel quantization from feedback at rate $\frac{\alpha}{2}\log(P)$,  then the total DoF are bounded above by $1+\alpha$. Furthermore, with quantized feedback of rate $\frac{\alpha}{2}\log(P)$ this DoF bound is achievable.

Finally, we go beyond 2 users and generalize the result to the $K$ user MISO broadcast channel  where the transmitter has $K$ antennas and there are $K$ users with a single antenna each. Here also, we prove that the DoF collapse to unity under non-degenerate channel uncertainty. Since the outer bound for this MISO BC is also an outer bound for MISO BC's with fewer than $K$ antennas at the transmitter or fewer than $K$ users, for $K$ user interference networks,  for $M\times N$ X channels where $K\geq\max(M,N)$, our result establishes the collapse of DoF to unity for all such networks under non-degenerate channel uncertainty. Remarkably, the best known outer bounds for $K$ user interference and $M\times N$ user $X$ networks under non-degenerate channel uncertainty (except for essentially degraded settings) prior to this work were $\frac{K}{2}$ and $\frac{MN}{M+N-1}$ (same as with perfect CSIT). Thus, at least for DoF, this work represents a pessimistic leap of the same magnitude as the ones made in the optimistic direction in \cite{Cadambe_Jafar_int, Cadambe_Jafar_X}.

\subsection{Notation}  We use the Landau $O(\cdot)$, $o(\cdot)$, and $\Theta(\cdot)$ notations as follows. For  functions $f(x), g(x)$ from $\mathbb{R}$ to $\mathbb{R}$, $f(x)=O(g(x))$ denotes that $\limsup_{x\rightarrow\infty}\frac{|f(x)|}{|g(x)|}<\infty$.  $f(x)=o(g(x))$ denotes that $\limsup_{x\rightarrow\infty}\frac{|f(x)|}{|g(x)|}=0$. $f(x)=\Theta(g(x))$ denotes that there exists a positive finite constant, $M$,  such that $\frac{1}{M} g(x)\leq f(x)\leq Mg(x)$, $\forall x$. We use $\mathbb{P}(\cdot)$ to denote the probability function $\mbox{Prob}(\cdot)$. We define $\lfloor x\rfloor$ as the largest integer that is smaller than or equal to $x$ when $x>0$,  the smallest integer that is larger than or equal to $x$ when $x<0$, and $x$ itself when $x$ is an integer. The index set $\{1,2,\cdots, n\}$ is represented compactly as $[1:n]$ or simply $[n]$ when it would cause no confusion. Arbitrary subsets of $[n]$ may be denoted as $[s]\subset[n]$. The difference of sets $[n]/[s]$ represents the set of elements that are in $[n]$ but not in $[s]$. $X^{[s]}$ represents $\{X(t): t\in[s]\}$. For example, $X^{[n]}=\{X(1), X(2),\cdots, X(n)\}$. With some abuse of notation we use $\{X\}$ to denote the set of values that can be taken by the random variable $X$. The cardinality of a set $A$ is denoted as $|A|$.

\section{The 2 User MISO BC with Perfect CSIT for  One User}
To prove the collapse of DoF in the strongest sense possible, let us first enhance the 2 user MISO BC by allowing perfect CSIT for user 1. Consider the vector broadcast channel with 2 users where the transmitter is equipped with $2$ antennas,  each user is equipped with $1$ receive antenna, and there are $2$ independent messages $W_1,W_2$ that originate at the transmitter and are desired by users $1$ and $2$, respectively. The transmission takes place over $n$ channel uses. The channel state information at the transmitter (CSIT) is denoted as $\mathcal{T}$, and includes perfect channel state information  for the channel vector of user  but not for the channel vector of user 2. In the terminology of Tandon et al. \cite{Tandon_Jafar_Shamai_Poor}, this is the PN setting,  although not restricted to any statistical equivalence assumptions.

The best outer bound for the DoF of the PN setting based on known results so far is $\frac{3}{2}$, which is  obtained from the finite state compound model by Weingarten, Shamai and Kramer in \cite{Weingarten_Shamai_Kramer} and is applicable to finite precision CSIT as well. While Weingarten et al. conjectured that their outer bound was loose even in the finite state compound setting, predicting a collapse of DoF, this conjecture was shown to be false by Gou, Jafar and Wang in \cite{Gou_Jafar_Wang}, who showed that $\frac{3}{2}$ DoF are achievable under the finite state compound model, through the DoF tuple $(d_1, d_2)=(1,0.5)$. The key to achievability is to split user 1's 1 DoF into two parts that carry 0.5 DoF each. These parts align at user 2, consuming half the available signal space of user 2, while remaining resolvable at user 1. User 2's signal, carrying 0.5 DoF, is then sent in the null space of user 1's channel, and is resolvable from the 0.5 dimensional interference-free space at user 2. Note that zero forcing at user 1 is possible because perfect CSIT for user 1 is assumed to be available. 

The $\frac{3}{2}$ DoF outer bound is also applicable in the blind interference alignment setting (BIA) introduced by Jafar in \cite{Jafar_corr}, where user 1 experiences time or frequency selective fading but user 2 experiences a relatively flat fading channel. Here also the outer bound is shown to be achievable  through the tuple $(d_1,d_2)=(1, 0.5)$. The key is to send two symbols for user 1, one from each antenna, repeated over two channel realizations where the channel of user 1 changes but the channel of user 2 remains the same. Thus, user 1 sees two linear combinations of the two symbols from which both symbols can be resolved, whereas user 2 only sees the same linear combination over both channel uses. Thus the interference occupies only 0.5 DoF at user 2. The remaining 0.5 DoF at user 2 is utilized by sending his desired signal, carrying 0.5 DoF, into the null space of user 1.

The finite state compound setting and the blind interference alignment setting reveal some of the challenges of proving the collapse of DoF for the PN setting. Any attempt at proving a collapse of DoF must carefully exclude such scenarios from the channel model. With this cautionary note, we are now ready to introduce the channel model for our problem.

\subsection{General Channel Model}
 The channel is described as follows:
\begin{eqnarray}
\left[\begin{array}{l}
\tilde{Y}_1(t)\\\tilde{Y}_2(t)
\end{array}\right]
&=&\underbrace{\left[\begin{array}{ll}
\tilde{G}_{11}(t)&\tilde{G}_{12}(t)\\
\tilde{G}_{21}(t)&\tilde{G}_{22}(t)
\end{array}\right]}_{{\bf \tilde{G}}(t)}
\left[\begin{array}{l}
\tilde{X}_1(t)\\
\tilde{X}_2(t)
\end{array}\right]+\left[\begin{array}{l}
\tilde{Z}_1(t)\\\tilde{Z}_2(t)
\end{array}\right]\label{eq:channelmodel}
\end{eqnarray}
where all symbols are real. At time $t\in\mathbb{N}$, $\tilde{Y}_k(t)$ is the  symbol received by user $k$, $\tilde{Z}_k(t)\sim\mathcal{N}(0,1)$ is the  real additive white Gaussian noise (AWGN), $\tilde{X}_k(t)$ is the  symbol sent from transmit antenna $k$, and $\tilde{G}_{kj}(t)$ is the channel fading coefficient between the  $j^{th}$ transmit antenna and user $k$.  The channel coefficients are not restricted to i.i.d. realizations, i.e., they may be correlated across space and time but are assumed to be drawn from a continuous distribution such that their joint density exists. The transmitter is subject to the  power constraint:
\begin{eqnarray}
\frac{1}{n}\sum_{t=1}^n[(\tilde{X}_1(t))^2+(\tilde{X}_2(t))^2]&\leq&\tilde{P}=O(P), \label{eq:power}
\end{eqnarray}

To avoid degenerate situations we will assume that the range of values of each of the elements $\tilde{G}_{ij}$ is bounded away from zero and infinity, as is the determinant of the overall channel matrix --- i.e., $|\tilde{G}_{ij}(t)|, \mbox{det}(\tilde{\bf G}(t))$ are all $\Theta(1)$. Stated explicitly, there exists positive finite constant $M$, such that
\begin{eqnarray}
\frac{1}{M}\leq |\tilde{G}_{ij}(t)|, \mbox{det}(\tilde{\bf G}(t))\leq M
\end{eqnarray}

 Note that this is not a major restriction because by choosing the bounding constants large enough, the omitted neighborhoods can be reduced to a probability measure less than $\epsilon$ for arbitrarily small $\epsilon$, and thus has only a vanishing impact on the DoF.

\subsection{Canonical Form}
Without loss of generality, for the purpose of deriving a DoF outer bound the channel model is reduced to the following  form, which is preferable due to the consolidation of channel parameters (See Appendix \ref{app:z} for justification). This is the canonical form that we will use throughout the paper.

\begin{figure}[h]
\center
\includegraphics[width=9cm]{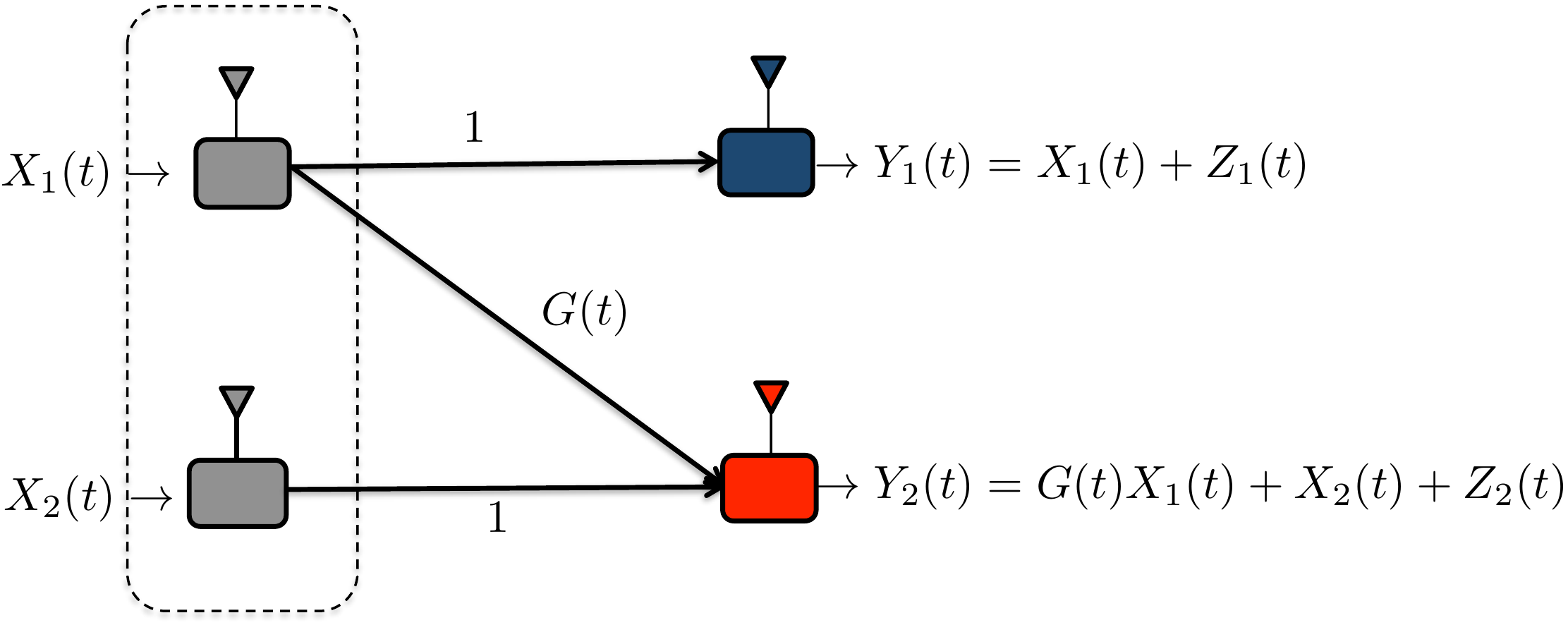}
\caption{Canonical form of the 2 user MISO BC with perfect CSIT for user 1. }\label{fig:canonicalbc}
\end{figure}
The canonical form of the channel model, shown in Fig. \ref{fig:canonicalbc} has the same  outputs $Y_1(t), Y_2(t)\in\mathbb{R}$, but the inputs are $X_1(t), X_2(t)\in\mathbb{R}$, so that:
\begin{eqnarray}
\left[\begin{array}{l}
Y_1(t)\\Y_2(t)
\end{array}\right]
&=&\left[\begin{array}{cc}
1&0\\
{G}(t)&1
\end{array}\right]
\left[\begin{array}{l}
{X}_1(t)\\
{X}_2(t)
\end{array}\right]+\left[\begin{array}{l}
Z_1(t)\\Z_2(t)
\end{array}\right]\label{eq:canonical}
\end{eqnarray}
The channel coefficient $G(t)$ is also bounded away from zero and infinity, i.e., there exists finite positive $M$, such that $|G(t)|\in\left(\frac{1}{M}, M\right)$.
The new power constraint is expressed as
\begin{eqnarray}
\frac{1}{n}\sum_{t=1}^n[({X}_1(t))^2+({X}_2(t))^2]&\leq&{P}, \label{eq:powercanonical}
\end{eqnarray}
where $P=\Theta(\tilde{P})$. Further, for notational convenience let us define the set of admissible inputs.
\begin{eqnarray}
\mathcal{X}^{[n]}&\triangleq&\{(X_1^{[n]},X_2^{[n]}): \frac{1}{n}\sum_{t=1}^n[({X}_1(t))^2+({X}_2(t))^2]\leq{P}\}
\end{eqnarray}

\subsection{Messages, Rates, Capacity, DoF}
The messages $W_1, W_2$ are jointly encoded at the transmitter for transmission over $n$ channel uses at rates $R_1, R_2$, respectively, into  a $2^{nR_1+ nR_2}\times n$ codebook matrix over the input alphabet. The codebook is denoted by $C(n,[R_1,R_2], P)$. For given power constraint parameter $P$, the rate vector $[R_1, R_2]$ is said to be achievable if there exists a sequence of codebooks $\mathcal{C}(n,[R_1,R_2], P)$, indexed by $n$, such that the probability that all messages are correctly decoded by their desired receivers approaches 1 as $n$ approaches infinity. The closure of achievable rate vectors is the capacity region $\mathcal{C}(P)$. The DoF tuple $(d_1, d_2)$ is said to be achievable if there exist $(R_1(P), R_2(P))\in\mathcal{C}(P)$ such that
\begin{eqnarray*}
d_1&=&\lim_{P\rightarrow\infty} \frac{R_1(P)}{\frac{1}{2}\log(P)}\\
d_2&=&\lim_{P\rightarrow\infty} \frac{R_2(P)}{\frac{1}{2}\log(P)}\\
\end{eqnarray*}
The closure of all achievable DoF tuples $(d_1, d_2)$ is called the DoF region, $\mathcal{D}$. The sum-DoF value is defined as 
\begin{eqnarray*}
\mathcal{D}_\Sigma&=&\max_{(d_1,d_2)\in\mathcal{D}}(d_1+d_2)
\end{eqnarray*}

\subsection{Non-degenerate Channel Uncertainty}
Beyond the assumptions that are already stated, the non-degenerate channel uncertainty model is defined by the additional assumption that the joint probability density function of the  channel coefficients, $G^{[n]}$, conditioned on the available CSIT, $\mathcal{T}$, which we denote as $f_{G^{[n]}|\mathcal{T}}(g^{[n]})$, exists and is bounded. Specifically, the conditions  that we require are the following.
\subsubsection{Peak of Density Function is Bounded for Fixed $P$}
For a given $P$, there exists a finite constant $f_{\max}$, $1\leq f_{\max}<\infty$, such that the probability that a subset of channel coefficients takes values in any measurable set is no more than the $\left(f_{\max}\right)^{|[s]|}$ times the Lebesgue measure of that set, where $|[s]|$ is the dimension of the space containing the set, i.e., the set is drawn from $\mathbb{R}^{|[s]|}$.
\begin{eqnarray}
 \mathbb{P}(G^{[s]}\in\mathcal{G}^{[s]}_o)&\leq&f_{\max}^{|[s]|}(P)\int_{\mathcal{G}^{[s]}_0}dG^{[s]}, ~~\forall [s]\subset[n], \forall \mathcal{G}^{[s]}_o\subset\mathcal{G}^{[s]}\label{eq:ourcondition}
\end{eqnarray}
The condition implies that a zero measure space cannot carry a non-zero probability. 
So it precludes scenarios where e.g., the channel is perfectly known or when one channel coefficient is a function of the rest. In all such cases, a zero measure space carries a non-zero probability, thus precluding the existence of a bounded constant $f_{\max}(P)$ as defined above. This restriction essentially accomplishes the same goal as the restriction by Lapidoth et al. \cite{Lapidoth_Shamai_Wigger_BC} that the differential entropy should be greater than $-\infty$. 

Note that if all joint and conditional density functions are bounded then (\ref{eq:ourcondition}) is immediately satisfied, and $f_{\max}$ may be chosen to be the peak value of all joint and conditional  density functions. 

\begin{eqnarray}
f_{\max}=\max\left(1,\max_{{[s]: [s]\subset[n]}}\sup_{\mathcal{G}^{[s]}} \left(f_{G^{[s]}|\mathcal{T},G^{[n]/[s]}}(g^{[s]})\right)^{\frac{1}{|[s]|}}\right)
\end{eqnarray}

Further, while we do not restrict channel realizations to be independent, suppose, just as an illustrative example, that we consider such a setting, which is of some interest. Then  the joint/conditional density   is  the $|[s]|$-fold product of the marginal density functions,
\begin{eqnarray}
f_{G^{[s]}|\mathcal{T}, G^{[n]/[s]}}(g^{[s]})&=&\prod_{t\in[s]} f_{G(t)|\mathcal{T}}(g(t))
\end{eqnarray}
and one can simply choose:
\begin{eqnarray}
f_{\max}&=&\max\left(1,\max_{t\in[1:n]}\sup_{g(t)\in\mathcal{G}(t)}f_{G(t)|\mathcal{T}}(g(t))\right)
\end{eqnarray}
Thus, a simple way to interpret our condition is that we require that the densities exist and are bounded.

\subsubsection{Peak of Density Function is Allowed to Scale with $P$}
\noindent As a function of $P$, we allow $f_{\max}(P)$ to scale as $O((\sqrt{P})^\alpha)$ for some $\alpha\in[0,1]$. 
 \begin{eqnarray}
 f_{\max}(P)&=&O(P^{\frac{\alpha}{2}})
 \end{eqnarray}
The case studied by Lapidoth et al. in \cite{Lapidoth_Shamai_Wigger_BC}, where the density does not depend on $P$, is represented here by setting $\alpha=0$. The positive values of $\alpha$ allow us to address settings where the CSIT improves with $P$, e.g., due to quantized channel feedback of rate $\frac{\alpha}{2}\log(P)$, so that the weight of the distribution is increasingly concentrated around the true channel realizations. Note that the maximum value of $\alpha$ is unity, because a feedback rate of $ \frac{1}{2}\log(P)$, implying 1 real DoF worth of feedback,  is sufficient to approach perfect CSIT performance over  channels that take only real values.

Since the receivers have full channel state information, $\mathcal{T}$ is globally known. For compact notation, we will suppress the conditioning, writing $f_{G^{[n]}}(g^{[n]})$ directly instead.

\subsection{$K$ User Extension}
Extending beyond the 2 user case, the canonical channel model in the $K$ user setting is described as follows.
\begin{figure}[h]
\center
\includegraphics[width=12cm]{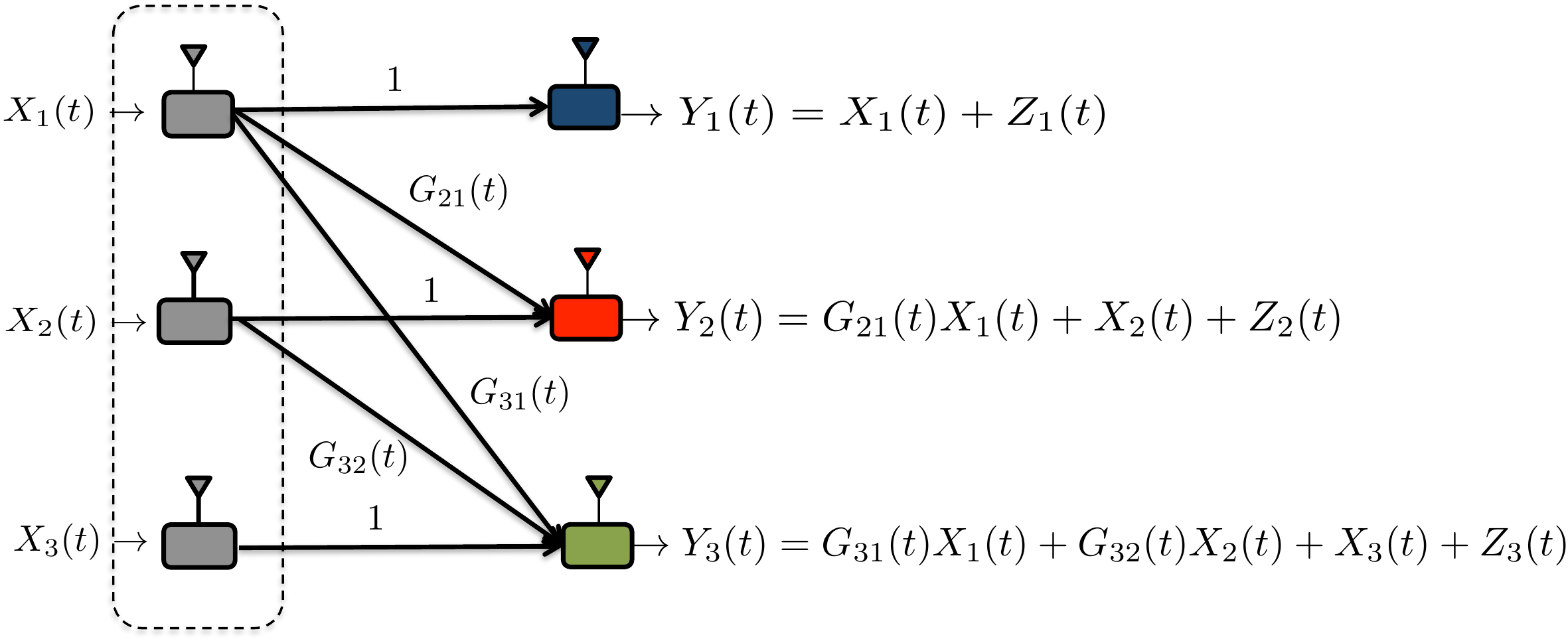}
\caption{Canonical form of the 3 user MISO BC with Graded Channel Uncertainty. }\label{fig:canonicalbc3}
\end{figure}\begin{eqnarray}
Y_1(t)&=&X_1(t)+Z_1(t)\\
Y_2(t)&=&G_{21}(t)X_1(t)+X_2(t)+Z_2(t)\\
&\vdots&\nonumber\\
Y_K(t)&=&G_{K1}(t)X_1(t)+G_{K2}(t)X_2(t)+\cdots+G_{K(K-1)}(t)X_{K-1}(t)+X_K(t)+Z_K(t)
\end{eqnarray}
where the inputs, $X_k(t)\in\mathbb{R}$, are subject to the power constraint
\begin{eqnarray}
\frac{1}{n}\sum_{t=1}^n\left((X_1(t))^2+(X_2(t))^2+\cdots+(X_K(t))^2\right)&\leq& P
\end{eqnarray}
The $G_{ij}(t)$ terms are  known to the transmitter only up to finite precision and are assumed to be bounded away from 0 and infinity. Further, the density of the $k^{th}$ users' unknown channel coefficients, $k>1$, is bounded by $f_{\max,k}(P)=O\left(P^{\frac{\alpha_k}{2}}\right)$. 

\section{Results}
We state the main result in its most general form, for $K$ users. The $2$ user case, corresponds to setting $\alpha_2\triangleq\alpha$.
\begin{theorem}\label{theorem:main}
For the $K$ user MISO BC with  non-degenerate channel uncertainty, the sum-GDoF are bounded above as
\begin{eqnarray}
\mathcal{D}_\Sigma&\leq&1+\alpha_2+\alpha_3+\cdots+\alpha_K
\end{eqnarray}
\end{theorem}
\subsubsection*{Settling the Conjecture by Lapidoth et al. in \cite{Lapidoth_Shamai_Wigger_BC}}
The $2$ user setting studied by Lapidoth et al., where the joint pdf is fixed, i.e., it does not depend on $P$, is captured here when $\alpha=0$ (equivalently, $\alpha_2=0)$. When $\alpha=0$,  the sum-GDoF are bounded above by unity, thus settling the conjecture of Lapidoth et al. for all non-degenerate channel uncertainty models. 

\subsubsection*{Settling the ``PN" Conjecture}
Since we allow perfect CSIT for one user, and one may assume (as a special case of our result) that the channels are i.i.d., the collapse of DoF for $\alpha=0$, also proves the conjecture of Tandon et al. for the 2 user setting.

\subsubsection*{Interference and $X$ networks}
Consider any one-hop wireless network where all receivers are equipped with a single antenna each. This includes all interference and $X$ networks. Allowing the transmitters to cooperate produces a MISO BC setting. Since  cooperation cannot hurt, the outer bound for the MISO BC  under non-degenerate channel uncertainty applies to interference and $X$ networks as well. In all cases, the DoF collapse to unity.

\subsubsection*{Limited rate feedback ($\alpha>0$)}
Consider the 2 user setting, with $\alpha=\alpha_2>0$. This case is  interesting because it has direct implications to the achievable DoF under limited rate quantized channel state feedback for the channel vector of user 2. If the feedback link has $\alpha$ DoF, i.e., the feedback rate scales as $\frac{\alpha}{2}\log(P)$ bits per channel use, then this corresponds to $\sim P^{\frac{\alpha}{2}}$ channel quantization levels, so that the size of a quantization interval scales as $\frac{1}{(\sqrt{P})^{\alpha}}$ and the channel density restricted to a quantization interval, i.e., $f_{G^{[n]}}(g^{[n]}|\mathcal{T})$ scales as $P^{\frac{\alpha}{2}}$. Theorem \ref{theorem:main} tells us that in this case the GDoF are bounded above as $\mathcal{D}_\Sigma\leq 1+\alpha$. It is also easy to see that under such quantized feedback, the DoF tuple $(d_1, d_2)=(1, \alpha)$ is achievable, simply by best-effort zero-forcing at the transmitter and treating residual interference as noise at the receiver 2. Thus, $\mathcal{D}_\Sigma=1+\alpha$ is the optimal sum-DoF value if the quantized channel state feedback is limited to rate $\frac{\alpha}{2}\log(P)$.  This  generalizes the result of Caire, Jindal and Shamai from \cite{Caire_Jindal_Shamai} who showed\footnote{While the result  in  \cite{Caire_Jindal_Shamai} is for the complex setting, the statement here is specialized to the real setting.} that in order to achieve the same DoF as with perfect CSIT, i.e., $\mathcal{D}_\Sigma=2$, the quantized feedback rate should scale as $\frac{1}{2}\log(P)$, i.e., carry one full degree of freedom $(\alpha=1)$. The bound for the $K$ user extension is similarly tight as well.

\allowdisplaybreaks
\section{Aligned Image Sets under Channel Uncertainty}
We begin this section with a disclaimer, that this section  is included mainly to share an intuitive understanding of the proof. The proof itself will be presented in  Section \ref{sec:proof2}. To facilitate an intuitive discussion, no attempt is made to be mathematically precise or rigorous here. The main idea we want to illustrate intuitively is a geometrical  notion of aligned images of codewords---loosely related to Korner and Marton's work on the images of a set in \cite{Korner_Marton_images} but under a much more specialized setting---which is the key to our proof.  As the proof in Section \ref{sec:proof2} will show, the problem boils down to the difference of two terms when only information to user 1 is being transmitted,
\begin{eqnarray}
\mathcal{D}_\Sigma &\leq& 1 + \limsup_{P\rightarrow\infty}\limsup_{n\rightarrow\infty}\frac{1}{\frac{n}{2}\log(P)}\left(h(Y_1^{[n]}|G^{[n]})-h(Y_2^{[n]}|G^{[n]})\right)\label{eq:start}
\end{eqnarray}
The first term, $h(Y_1^{[n]}|G^{[n]})$, we wish to maximize because it represents the rate of desired information   being sent to user 1. The second, $h(Y_2^{[n]}|G^{[n]})=h(G^{[n]}X_1^{[n]}+X_2^{[n]}+Z_2^{[n]}|G^{[n]})$ we wish to minimize, because it represents the interference seen by user 2, due to the information being sent to user 1. If $G^{[n]}$ was perfectly available to the transmitter, then $X_2^{[n]}$ could be chosen to cancel $G^{[n]}X_1^{[n]}$ thus eliminating interference entirely at user 2. With only statistical knowledge of $G^{[n]}$, zero forcing is not possible. Indeed, the purpose of $X_2^{[n]}$ is mainly to align interference into as small a space as possible. However, instead of consolidating interference in the sense of vector space dimensions, as is typically the case in DoF studies involving interference alignment, here the goal is for $X_2^{[n]}$ to minimize the size of the image, as seen by user 2,  of the codewords that carry information for user 1. This is the new perspective that is the key to the proof.

\subsection{Toy Setting to Introduce Aligned Image Sets}
For illustrative purposes, let us start with a rather extreme over-simplification, by considering the case with $n=1$, ignoring noise, and using the log of the cardinality of the codewords as a surrogate for the entropy. With this simplification, the quantity that we are interested in is the difference:
\begin{eqnarray}
\log|\{X_1\}|-\log|\{GX_1+X_2\}|
\end{eqnarray}
averaged over $G$. By $|\{A\}|$ is meant the cardinality of the set of values taken by the variable $A$.

The codebook is the set of ($X_1, X_2$) values. Note that $|\{X_1\}|$, the number of distinct values of $X_1$, is the number of distinct ``codewords"  as seen by user 1, who (once noise is ignored) only sees $Y_1=X_1$, so that his ``rate" is $\log|\{X_1\}|$. Given the set of  $X_1$ values, we would like to associate each $X_1$ value with a corresponding $X_2$ value, such that the number of distinct values of $Y_2=GX_1+X_2$ is minimized. In other words, we wish to minimize the image of the set of codewords as seen by user 2, by choosing $X_2$ to be a suitable function of $X_1$.

\begin{figure}[h]
\center
\includegraphics[width=7cm]{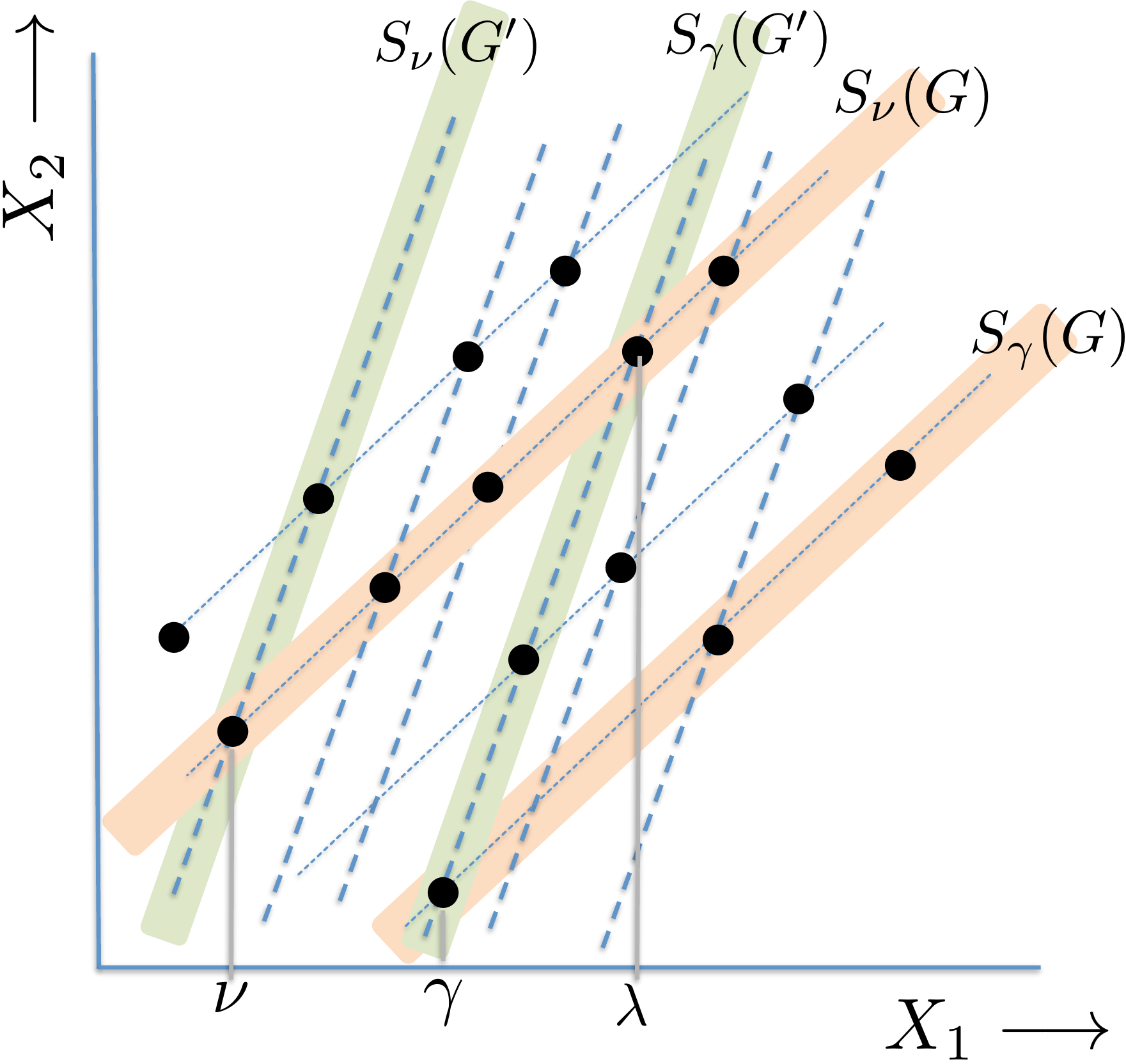}
\caption{Two codewords, $\nu$ and $\gamma$, and their equivalence classes, $S_\nu$ and $S_\gamma$, containing all codewords that have the same image at user 2 as $\nu$ and $\gamma$, respectively. The partitioning into equivalence classes depends on the channel realization. The figure shows the distinct equivalence classes for  two channel realizations, $G$ and $G'$. }\label{fig:eqclass}
\end{figure}

Consider two codewords $(X_1, X_2)=(x_1, x_2)$ and $(X_1,X_2)=(x_1',x_2')$. If $x_1\neq x_1'$ then these codewords are distinct from user 1's perspective, and thus capable of carrying information to user 1 via the transmitter's choice to transmit one or the other. Suppose the channel is $G$. Then for these two codewords to ``\emph{align}" where they cause interference, they must have the same image as seen by user 2. This gives us the  condition for aligned images that is central to this work.
\begin{eqnarray}
Gx_1+x_2&=&Gx_1'+x_2'\\
\Rightarrow G&=&-\left(\frac{x_2'-x_2}{x_1'-x_1}\right)
\end{eqnarray}
In other words, $G$ must be the negative of the slope of the line connecting the codeword  $(x_1,x_2)$ to the codeword  $(x_1',x_2')$ in the $X_1,X_2$ plane.  For a given channel realization $G$, all codewords that align with $(x_1,x_2)$  (i.e., whose images align with the image of $(x_1,x_2)$) as seen by user 2, must  lie on the same line that passes through $(x_1, x_2)$ and has slope $-G$. Conversely, all codewords that lie on this line have images that align with the image of $(x_1,x_2)$ at user 2. For any codeword that does not lie on this line, there is a parallel line with the same slope, $-G$, that represents the set of codewords whose images align with the image of that codeword. Thus, these lines of the same slope, $-G$, partition the set of codewords into  equivalence classes, such that codewords that lie on the same line have the same image at user 2. 
Also note that a different channel realization, $G'$, gives rise to a different equivalent class partition, corresponding to lines with slope $-G'$. This is illustrated in Fig. \ref{fig:eqclass}. Since the $X_2$ values are functions of $X_1$ values, in the figure we label the codewords only on the $X_1$ axis. The codeword $\nu$ belongs to the equivalence class $S_\nu(G)$ under the channel realization $G$ and to the equivalence class $S_\nu(G')$ under the channel realization $G'$. Also, note that two codewords that belong to the same equivalence class under one channel realization, cannot belong to the same equivalence class under any other channel realization. For instance, codewords $\lambda$ and $\nu$ belong to the same equivalence class $S_\nu(G)$ under channel realization $G$, but they belong to different equivalence classes, $S_\nu(G')$ and $S_\gamma(G')$, under a different channel realization $G'$. 

Recall that our goal is to minimize the number of distinct images seen by user 2, (averaged) across all channel realizations. It is not preferable to have too few codewords aligned into the same image, because it would create too many distinct images across all codewords. However, remarkably, it is also not preferable to have too many codewords aligned into the same image for any given channel realization, because for every \emph{other} channel realization, each of these codewords must have a different image. Thus,
\begin{eqnarray}
\mbox{No. of distinct images}&\geq&\mbox{No. of codewords per image}
\end{eqnarray}

A rough calculation (assuming uniformity for this intuitive argument so averages can be ignored) will shed light on the best case scenario. 
\begin{eqnarray}
\mbox{No. of codewords} &=& \mbox{No. of distinct images }\times\mbox{ No. of codewords per image}\\
&\leq& \mbox{No. of distinct images }\times\mbox{ No. of distinct images}\\
\Rightarrow \mbox{No. of distinct images} &\geq& \sqrt{\mbox{No. of codewords}}
\end{eqnarray}
Mathematically,
\begin{eqnarray}
|\{GX_1+X_2\}|&\geq&\sqrt{|\{X_1\}|}
\end{eqnarray}
and therefore we have the bound
\begin{eqnarray}
\log(|\{X_1\}|) - \log(|\{GX_1+X_2\}|)&\leq&\log(|\{X_1\}|)-\frac{1}{2}\log(|\{X_1\}|)=\frac{1}{2}\log(|\{X_1\}|)
\end{eqnarray}
Now, realizing that $\log(|\{X_1\}|)$ can at most be user 1's single interference free capacity, i.e., that it corresponds to at most rate $\frac{1}{2}\log(P)$, and substituting back into (\ref{eq:start}) the intuitive argument leads us to $\mathcal{D}_\Sigma \leq \frac{3}{2}$. This is indeed the DoF outer bound for the finite state compound channel setting, and is, quite surprisingly achievable, as shown by Gou, Jafar, Wang in \cite{Gou_Jafar_Wang} and Maddah-Ali in \cite{Maddah_Compound} when the number of possible channel realizations, $|\mathcal{G}|$, is finite and fixed, even as the cardinality of codebooks approaches infinity as  $P\rightarrow\infty$.

\subsection{Sketch of Proof}
Unfortunately, the utility of the over-simplified noise free setting with $n=1$ starts wearing thin as we go beyond this point. While the picture of aligned image sets illustrated here will continue to be meaningful, the remaining nuances of the proof are not easily conveyed in this toy setting. Staying with the intuitive character of this section, let us conclude with an outline which will be useful to navigate the structure of the proof that appears in the subsequent section. 

From the perspective of DoF studies, the presence of noise essentially imposes a resolution threshold, e.g., $\delta$, such that the codewords with images that differ by less than $\delta$, are unresolvable. As the first step of the proof, this effect is captured by discretizing the input and output alphabet and eliminating noise, as is done in a variety of deterministic channel models that have been used for DoF studies \cite{Avestimehr_Diggavi_Tse, Bresler_Tse}, so that instead of differential entropies we now need to deal only with entropies $H(\bar{Y}_1^{[n]}|G^{[n]})$ and $H(\bar{Y}_2^{[n]}|G^{[n]})$. Here $\bar{X}_1, \bar{X}_2$ represent the discretized inputs, $\bar{Y}_1, \bar{Y}_2$ the discretized outputs, and $\bar{Y}_1=\bar{X}_1$.  Next step is to note that we are only interested in the maximum value of the difference $H(\bar{Y}_1^{[n]}|G^{[n]})-H(\bar{Y}_2^{[n]}|G^{[n]})$. It then follows that without loss of generality, $\bar{X}_2^{[n]}$ can be made a function of $\bar{X}_1^{[n]}$, and therefore $\bar{Y}_2^{[n]}$ becomes a function of $\bar{Y}_1^{[n]}, G^{[n]}$. This implies that $H(\bar{Y}_1^{[n]}|G^{[n]})=H(\bar{Y}_1^{[n]}, \bar{Y}_2^{[n]}|G^n)$ $=H(\bar{Y}_2^{[n]}|G^{[n]})+H(\bar{Y}_1^{[n]}|\bar{Y}_2^{[n]},G^{[n]})$. Thus, the difference of entropies is equal to $H(\bar{Y}_1^{[n]}|\bar{Y}_2^{[n]},G^{[n]})=H(\bar{X}_1^{[n]}|\bar{Y}_2^{[n]},G^{[n]})$. Now, conditioned on $\bar{Y}_2^{[n]}, G^{[n]}$, the set of feasible values of $\bar{X}_1^{[n]}$ is precisely an aligned image set $S(G^{[n]})$, i.e.,  all these $\bar{X}_1^{[n]}$  produce the same value of $\bar{Y}_2^{[n]}$ for the given channel realization $G^{[n]}$. Since entropy is maximized by a uniform distribution, $H(\bar{X}_1^{[n]}|\bar{Y}_2^{[n]},G^{[n]})\leq \mbox{E}_{G^n}\left[\log\left(|S(G^{[n]})|\right)\right]\leq\log\left(\mbox{E}|S(G^{[n]})|\right)$, where the last step followed from Jensen's inequality. Thus, the difference of entropies is bounded by the log of the expected cardinality of the aligned image sets. The most critical step of the proof then is to bound the expected cardinality of aligned image sets. This is done by bounding the probability that two given $\bar{X}_1^{[n]}$ are in the same aligned image set, i.e., the probability of the set of channels for which the two produce the same image $\bar{Y}_2^{[n]}$. Recall that for two codewords to belong to the same aligned set in the absence of noise, the channel realization over each channel use must be the slope of the vector connecting the corresponding codeword vectors. The  blurring of  $\delta$ around the two codewords also blurs the slope of the line connecting them, but by no more than $\pm \delta/\Delta$, where $\Delta$ is the distance (difference in magnitudes) between the two codeword symbols over that channel use. Thus, the probability that the given two codewords that are resolvable at user 1 cast the same image at user 2 is bounded above by $\approx f_{\max}\frac{2\delta}{\Delta}$. The power constraint of $P$ implies that there are at most $\approx \sqrt{P}/{\delta}$ resolvable codeword symbols per channel use. Summing over all possible resolvable codeword symbols, gives us  $\approx f_{\max}\sum_{\Delta\in[0:\sqrt{P}/\delta]}\frac{2\delta}{\Delta}= f_{\max}\log(P)+o(\log(P))$, per channel use, so that the average cardinality of an aligned image set, $E|S(G^{[n]}|$, turns out to be bounded above by $\approx(f_{\max}\log(P))^n$, and $\log(E|S(G^{[n]}|)$ is bounded above by $\approx n\log(f_{\max})+n\log(\log(P))$. Since $f_{\max}=O(P^{\frac{\alpha}{2}})$, normalizing by $\frac{n}{2}\log(P)$ and sending first $n$ and then $P$ to infinity sends this term to $\alpha$. Thus, combining with (\ref{eq:start}) produces the sum-DoF outer bound value $1+\alpha$, giving us the result of Theorem 1. Note that in the DoF limit, $\delta=\Theta(1)$, and it will be useful to think of it as $1$ for simplicity, so that the inputs and outputs are restricted to integer values. With this sketch as the preamble, we now proceed to the actual proof.

\section{Proof of Theorem \ref{theorem:main} for $K=2$ Users }\label{sec:proof2}
For ease of exposition, the proof is divided into several key steps. The first step is the discretization of the channel to capture the effect of noise, leading to a deterministic channel model, whose DoF will be an outer bound to the DoF of the  canonical channel model, which in turn is an outer bound on the DoF of the  general channel model.
\begin{enumerate}
\item {\bf Deterministic Channel Model}\\
The  deterministic channel model has inputs $\bar{X}_1(t), \bar{X}_2(t)\in\mathbb{Z}$ and outputs $\bar{Y}_1(t),\bar{Y}_2(t)\in\mathbb{Z}$, defined as
\begin{eqnarray}
\bar{Y}_1(t)&=&\bar{X}_1(t)\\
\bar{Y}_2(t)&=&\lfloor G(t)\bar{X}_1(t)\rfloor+\bar{X}_2(t)
\end{eqnarray}
with the power constraint
\begin{eqnarray}
\bar{X}_1(t), \bar{X}_2(t)\in\{0,1,\cdots,\lceil\sqrt{P}\rceil\}, ~\forall t\in\mathbb{N}
\end{eqnarray}
and the set of inputs that satisfy the power constraints defined as
\begin{eqnarray}
\bar{\mathcal{X}}^{[n]}&=&\{(\bar{X}_1^{[n]},\bar{X}_2^{[n]})\in\mathbb{Z}^{[n]}\times\mathbb{Z}^{[n]} :  \bar{X}_1(t), \bar{X}_2(t)\in\{0,1,\cdots,\lceil \sqrt{P} \rceil\}, ~\forall t\in[1:n]\}
\end{eqnarray}
The assumptions on the unknown channel coefficients sequence $G^{[n]}$ are the same as before.
\begin{lemma}\label{lemma:detn}
The DoF of the canonical channel model are bounded above by the DoF of the deterministic channel model.
\end{lemma}
The proof of Lemma \ref{lemma:detn} appears in Appendix \ref{app:detn} and follows along the lines of similar proofs by Bresler and Tse in \cite{Bresler_Tse}.
\item{\bf Difference of Entropies Representing Desired Signal and Interference Dimensions}\\
Starting from Fano's inequality, we proceed as follows.
\begin{eqnarray}
nR_1&\leq&I(W_1; \bar{Y}_1^{[n]}|W_2, G^{[n]})+o(n)\label{eq:startn}\\
&=&H(\bar{Y}_1^{[n]}|W_2, G^{[n]})+o(n)\\
nR_2&\leq&I(W_2;\bar{Y}_2^{[n]}|G^{[n]})+o(n)\\
&=&H(\lfloor G^{[n]} \bar{X}_1^{[n]}\rfloor +\bar{X}_2^{[n]}|G^{[n]})-H(\bar{Y}_2^{[n]}|W_2, G^{[n]})+o(n)\\
&\leq & \frac{n}{2}\log(P)-H(\bar{Y}_2^{[n]}|W_2, G^{[n]})+n~o(\log(P))+o(n)\label{eq:onedof}\\
\Rightarrow n(R_1+R_2)&\leq&\frac{n}{2}\log(P)+[H(\bar{Y}_1^{[n]}|W_2,G^{[n]})-H(\bar{Y}_2^{[n]}|W_2,G^{[n]})]+n~o(\log(P))+o(n)\\
\Rightarrow \mathcal{D}_\Sigma&\leq&1+\limsup_{P\rightarrow\infty}\limsup_{n\rightarrow\infty}\frac{[H(\bar{Y}_1^{[n]}|W_2,G^{[n]})-H(\bar{Y}_2^{[n]}|W_2,G^{[n]})]}{\frac{n}{2}\log(P)}\\
&\leq&1+\limsup_{P\rightarrow\infty}\limsup_{n\rightarrow\infty}\max_{w_2\in[1:2^{nR_2}]}\frac{[H(\bar{Y}_1^{[n]}|W_2=w_2,G^{[n]})-H(\bar{Y}_2^{[n]}|W_2=w_2,G^{[n]})]}{\frac{n}{2}\log(P)}\nonumber\\ 
&=&1+\bar{\mathcal{D}}_\Delta\label{eq:Deltan}
\end{eqnarray}
so that what remains is to bound the difference of entropy terms:
\begin{eqnarray}
\bar{\mathcal{D}}_\Delta&\triangleq&\limsup_{P\rightarrow\infty}\limsup_{n\rightarrow\infty}\max_{\substack{\mathbb{P}(\bar{X}_1^{[n]},\bar{X}_2^{[n]})\\  (\bar{X}_1^{[n]},\bar{X}_2^{[n]})\in\bar{\mathcal{X}}^{[n]}}}\frac{H(\bar{X}_1^{[n]}|G^{[n]})-H(\lfloor G^{[n]}\bar{X}_1^{[n]}\rfloor+\bar{X}_2^{[n]}|G^{[n]})}{\frac{n}{2}\log(P)} \label{eq:detkey}
\end{eqnarray} 
Note that in (\ref{eq:onedof}) we bounded $H(\lfloor G^{[n]} \bar{X}_1^{[n]}\rfloor +\bar{X}_2^{[n]}|G^{[n]})$ as follows.
\begin{eqnarray}
H(\lfloor G^{[n]} \bar{X}_1^{[n]}\rfloor +\bar{X}_2^{[n]}|G^{[n]})&\leq&\sum_{t=1}^nH(\lfloor G(t) \bar{X}_1(t)\rfloor +\bar{X}_2(t)|G(t))\\
&\leq&\sum_{t=1}^n\mbox{E}\log\left((|G(t)|+1)(\lceil\sqrt{P}\rceil+1)\right)\label{eq:card}\\
&=&\frac{n}{2}\log(P)+n~o(\log(P))
\end{eqnarray}
where (\ref{eq:card}) follows from the observation that for a given $G(t)$ value, $\lfloor G^{[n]} \bar{X}_1^{[n]}\rfloor +\bar{X}_2^{[n]}$ can take at most $(1+G(t))(1+\lceil\sqrt{P}\rceil)$ values (all integers) and the entropy of a variable that can take finitely many values is at most the log of the number of values.

\item{\bf Functional Dependence $\bar{X}_2^{[n]}(\bar{X}_1^{[n]})$}\\
Next we show that one can assume that $\bar{X}_2^{[n]}$ is a function of $\bar{X}_1^{[n]}$. Given the sets of codeword vectors $\{\bar{X}_1^{[n]}\}$, $\{\bar{X}_2^{[n]}\}$, define $\mathcal{L}$ as the mapping from $\bar{X}_1^{[n]}$ to $\bar{X}_2^{[n]}$, i.e.,
\begin{eqnarray}
\bar{X}_2^{[n]}&=&\mathcal{L}(\bar{X}_1^{[n]})
\end{eqnarray} 
In general, because the mapping may be random, $\mathcal{L}$ is a random variable. Because conditioning cannot increase entropy,
\begin{eqnarray}
H\left(\lfloor G^{[n]}\bar{X}_1^{[n]}\rfloor+\mathcal{L}(\bar{X}_1^{[n]})\right|G^{[n]})&\geq&H\left(\lfloor G^{[n]}\bar{X}_1^{[n]}\rfloor+\mathcal{L}(\bar{X}_1^{[n]})\right|G^{[n]}, \mathcal{L})\\
&\geq&\min_{L\in\{\mathcal{L}\}}H\left(\lfloor G^{[n]}\bar{X}_1^{[n]}\rfloor+\mathcal{L}(\bar{X}_1^{[n]})\right|G^{[n]}, \mathcal{L}=L)
\end{eqnarray}
Let $L_o\in\mathcal{L}$ be the mapping that minimizes the entropy term. Then, choosing
\begin{eqnarray}
\bar{X}_2^{[n]}(\bar{X}_1^{[n]})&=&L_o(\bar{X}_1^{[n]})
\end{eqnarray}
we have 
\begin{eqnarray}
\bar{\mathcal{D}}_\Delta&\leq&\hat{\mathcal{D}}_\Delta\triangleq\limsup_{P\rightarrow\infty}\limsup_{n\rightarrow\infty}\max_{\substack{\mathbb{P}(\bar{X}_1^{[n]}), \bar{X}_2^{[n]}(\bar{X}_1^{[n]})\\  (\bar{X}_1^{[n]},\bar{X}_2^{[n]})\in\bar{\mathcal{X}}^{[n]}}}\frac{H(\bar{X}_1^{[n]}|G^{[n]})-H(\lfloor G^{[n]}\bar{X}_1^{[n]}\rfloor+\bar{X}_2^{[n]}(\bar{X}_1^{[n]})|G^{[n]})}{\frac{n}{2}\log(P)}\nonumber\\
&&\label{eq:Deltahatn}
\end{eqnarray}
because the choice of the mapping function does not affect the positive entropy term, and it minimizes the negative entropy term. Henceforth, because $\bar{X}_2^{[n]}$ is a function of $\bar{X}_1^{[n]}$, we will refer to codewords only through $\bar{X}_1^{[n]}$ values. 
\item {\bf Definition of Aligned Image Sets}\\
The aligned image set containing the codeword $\bar{\nu}^{[n]}\in\{\bar{X}_1^{[n]}\}$ for channel realization $G^{[n]}$ is defined as the set of all codewords that cast the same image as $\bar{\nu}^{[n]}$ at user 2. 
\begin{eqnarray}
S_{\bar{\nu}^{[n]}}(G^{[n]})&\triangleq&\{\bar{x}_1^{[n]}\in\{\bar{X}_1^{[n]}\}: \lfloor G^{[n]}\bar{x}_1^{[n]}\rfloor+\bar{X}_2^{[n]}(\bar{x}_1^{[n]})=\lfloor G^{[n]}\bar{\nu}^{[n]}\rfloor+\bar{X}_2^{[n]}(\bar{\nu}^{[n]})\}
\end{eqnarray}
Since we will need the average (over $G^{[n]}$) of the cardinality of an aligned image set, E$|S_{\bar{\nu^{[n]}}}(G^{[n]})|$, it is worthwhile to point out that the cardinality $|S_{\bar{\nu^{[n]}}}(G^{[n]})|$ as a function of $G^{[n]}$, is a bounded simple function, and therefore measurable. It is bounded because its values are restricted to natural numbers not greater than $(1+\lceil \sqrt{P}\rceil)^{2n}$. To see that it is a simple function,  note that $|S_{\bar{\nu}^{[n]}}(G^{[n]})|$ is continuous on irrational numbers. Therefore, the set $\{G^{[n]}: |S_{\bar{\nu}^{[n]}}(G^{[n]})|=m\}, \forall m\in \{1,\cdots,(1+\lceil \sqrt{P}\rceil)^{2n}\}$ is the union of two measurable sets. The first is an intersection of an open set and the set of irrational numbers, so it is measurable.  The second is a subset of rational numbers which generally has zero measure. So, $\{G^{[n]}: |S_{\bar{\nu}^{[n]}}(G^{[n]})|=m\}$ is a measurable set.

\item{\bf Bounding Difference of Entropies, $\hat{\mathcal{D}}_\Delta$, in Terms of Size of Aligned Image Sets}\\
\begin{eqnarray}
H(\bar{X}_1^{[n]}|G^{[n]})&=& H(\bar{X}_1^{[n]}, S_{\bar{X}_1^{[n]}}(G^{[n]})|G^{[n]})\\
&=&H(S_{\bar{X}_1^{[n]}}(G^{[n]})|G^{[n]})+H(\bar{X}_1^{[n]}| S_{\bar{X}_1^{[n]}}(G^{[n]}),G^{[n]})\\
&=&H(\lfloor G^{[n]}\bar{X}_1^{[n]}\rfloor+\bar{X}_2^{[n]}(\bar{X}_1^{[n]})|G^{[n]})+H(\bar{X}_1^{[n]}| S_{\bar{X}_1^{[n]}}(G^{[n]}),G^{[n]})\\
&\leq&H(\lfloor G^{[n]}\bar{X}_1^{[n]}\rfloor+\bar{X}_2^{[n]}(\bar{X}_1^{[n]})|G^{[n]})+\mbox{E}\left[\log(|S_{\bar{X}_1^{[n]}}(G^{[n]})|)\right]\label{eq:uniform}\\
&\leq&H(\lfloor G^{[n]}\bar{X}_1^{[n]}\rfloor+\bar{X}_2^{[n]}(\bar{X}_1^{[n]})|G^{[n]})+\log\left(\mbox{E}\left[|S_{\bar{X}_1^{[n]}}(G^{[n]})|\right]\right)\label{eq:jensensn}
\end{eqnarray}
where (\ref{eq:uniform}) follows because uniform distribution maximizes entropy, and (\ref{eq:jensensn}) follows from Jensen's inequality. Rearranging terms, we note that
\begin{eqnarray}
\hat{\mathcal{D}}_\Delta&\leq&\limsup_{P\rightarrow\infty}\limsup_{n\rightarrow\infty}\max_{\substack{P(\bar{X}_1^{[n]}), \bar{X}_2^{[n]}(\bar{X}_1^{[n]})\\  (\bar{X}_1^{[n]},\bar{X}_2^{[n]})\in\bar{\mathcal{X}}^{[n]}}}\frac{\log\left(\mbox{E}\left[|S_{\bar{X}_1^{[n]}}(G^{[n]})|\right]\right)}{\frac{n}{2}\log(P)}\label{eq:DS}
\end{eqnarray}

\item{\bf Bounding the Probability of Image Alignment}\\
Given two codewords $\bar{x}_1^{[n]}$ and $\bar{\nu}^{[n]}$, let us bound the probability that their images align at user 2. Note that for $\bar{x}_1^{[n]}\in S_{\bar{\nu}^{[n]}}(G^{[n]})$  we must have 
\begin{eqnarray}
\lfloor G^{[n]}\bar{x}_1^{[n]}\rfloor-\lfloor G^{[n]}\bar{\nu}^{[n]}\rfloor&=&\bar{X}_2^{[n]}(\bar{\nu}^{[n]})-\bar{X}_2^{[n]}(\bar{x}_1^{[n]})\\
\Rightarrow G^{[n]}(\bar{x}_1^{[n]}-\bar{\nu}^{[n]})&\in&\bar{X}_2^{[n]}(\bar{x}_1^{[n]})-\bar{X}_2^{[n]}(\bar{\nu}^{[n]})+\Delta_{(-1,1)}^{[n]}
\end{eqnarray}
where $\Delta_{(-1,1)}(t)\in(-1,1), \forall t\in[1:n]$. Thus, for all $t\in[1:n]$ such that $\bar{x}_1(t)\neq\bar{\nu}(t)$, the value of $G(t)$ must lie within an interval of length no more than $\frac{2}{|\bar{x}_1(t)-\bar{\nu}(t)|}$. Since the maximum value of the joint probability density function of $\{G(t):  \mbox{ such that } \bar{x}_1(t)\neq\bar{\nu}(t), t\in[1:n]\}$ is bounded by $f_{\max}^{\sum_{t=1}^n 1(\bar{x}_1(t)\neq\bar{\nu}(t))}\leq f_{\max}^{[n]}$, we  can bound the probability that the images of two codewords align as follows.
\begin{eqnarray}
\mathbb{P}(\bar{x}_1^{[n]}\in S_{\bar{\nu}^{[n]}}(G^{[n]}))&\leq&f_{\max}^{n}\prod_{t:\bar{x}_1(t)\neq\bar{\nu}(t)}\frac{2}{\left|\bar{x}_1(t)-\bar{\nu}(t)\right|}
\end{eqnarray}

\item{\bf Bounding the Average Size of  Aligned Image Sets}\\
\begin{eqnarray}
\mbox{E}\left[\left|S_{\bar{\nu}^{[n]}}(G^{[n]})\right|\right]&=&\sum_{\bar{x}_1^{[n]}\in\{\bar{X}_1^{[n]}\}}\mathbb{P}\left(\bar{x}_1^{[n]}\in S_{\bar{\nu}^{[n]}}(G^{[n]}) \right)\\
&=&1+\sum_{\substack{\bar{x}_1^{[n]}\in\{\bar{X}_1^{[n]}\}\\ \bar{x}_1^{[n]}\neq \bar{\nu}^{[n]}}}\mathbb{P}\left(\bar{x}_1^{[n]}\in S_{\bar{\nu}^{[n]}}(G^{[n]}) \right)\\
&\leq&1+(2f_{\max})^{n}\sum_{\substack{\bar{x}_1^{[n]}\in\{\bar{X}_1^{[n]}\}\\ \bar{x}_1^{[n]}\neq \bar{\nu}^{[n]}}}\prod_{t:\bar{x}_1(t)\neq\bar{\nu}(t)}\frac{1}{\left|\bar{x}_1(t)-\bar{\nu}(t)\right|}\\
&\leq&1+(2f_{\max})^{n}\prod_{t=1}^n\left(1+\sum_{\Delta_{\bar{x}}=1}^{\lceil\sqrt{P}\rceil}\frac{2}{\Delta_{\bar{x}}}\right)\\
&\leq&1+(2f_{\max})^{n}\prod_{t=1}^n\left(\log(\sqrt{P})+o(\log(P))\right)
\end{eqnarray}
Since this is true for all $\bar{\nu}^{[n]}\in\{\bar{X}_1^{[n]}\}$
\begin{eqnarray}
\mbox{E}\left[\left|S_{\bar{X}_1^{[n]}}(G^{[n]})\right|\right]&\leq&1+(2f_{\max})^{n}\left(\log(\sqrt{P})+o(\log(P))\right)^{n}\label{eq:avesizen}
\end{eqnarray}

\item{\bf Combining the Bounds to Complete the Proof}\\
Combining (\ref{eq:DS}) and (\ref{eq:avesizen}) we have
\begin{eqnarray}
\hat{\mathcal{D}}_\Delta&\leq&\limsup_{P\rightarrow\infty}\limsup_{n\rightarrow\infty}\frac{\log\left(1+(2f_{\max})^{n}\left(\log(\sqrt{P})+o(\log(P))\right)^{n} \right)}{\frac{n}{2}\log(P)}\\
&=&\limsup_{P\rightarrow\infty}\frac{\log(f_{\max})}{\frac{1}{2}\log(P)}+\limsup_{P\rightarrow\infty}\frac{\log(\log(P))}{\frac{1}{2}\log(P)}\\
&\leq&\alpha \label{eq:alphan}
\end{eqnarray}
where (\ref{eq:alphan}) follows because $f_{\max}=O(P^{\frac{\alpha}{2}})$. 
Finally combining (\ref{eq:alphan}) with (\ref{eq:Deltan}) and (\ref{eq:Deltahatn}) we have the desired outer bound
\begin{eqnarray}
\mathcal{D}_\Sigma&\leq&1+\alpha
\end{eqnarray}
\end{enumerate}
\hfill$\Box$

\section{Proof of Theorem \ref{theorem:main} for $K$ Users}\label{sec:proofk}
The generalization of the proof to the $K$ user setting is, for the most part, straightforward based on the 2 user case studied earlier. To avoid repetition our presentation will only briefly summarize the aspects that follow directly and use detailed exposition for only those aspects that require special attention. We divide the proof into a similar set of steps for ease of reference with the 2 user case.
\begin{enumerate}
\item {\bf Deterministic Channel Model}\\
As in the 2 user case, the deterministic channel model is described as:
\begin{eqnarray}
\bar{Y}_k&=&\sum_{i=1}^{k-1}\lfloor G_{ki}(t)\bar{X}_i(t)\rfloor +\bar{X}_k(t)
\end{eqnarray}
where the integer inputs satisfy the following per-symbol power constraint
\begin{eqnarray}
\bar{X}_k(t)&\in&\{0,1,\cdots, \lceil\sqrt{P}\rceil\}, ~~\forall k\in[1:K]
\end{eqnarray}
As before, let us define $\bar{\mathcal{X}}^{[n]}$ as the set of codewords that satisfy the power constraint. We have the following bound.
\begin{lemma}\label{lemma:kdet}
The DoF of the canonical model are bounded above by the DoF of the deterministic model.
\end{lemma}
We omit the proof of Lemma \ref{lemma:kdet} since it is a straightforward extension of the 2 user proof which was already presented in much detail.
\item {\bf Difference of Entropy Terms}\\
For the $k^{th}$ user we bound the rate as
\begin{eqnarray}
nR_k&\leq&I(W_k; \bar{Y}_k^{[n]}|G^n,W_{k+1}, W_{k+2}, \cdots, W_K)+o(n)\\
&\leq& H(\bar{Y}_k^{[n]}|G^n,W_{k+1}, \cdots, W_K)-H(\bar{Y}_k^{[n]}|G^n,W_k, W_{k+1}, \cdots, W_K)+o(n)
\end{eqnarray}
where  $G^n$ includes all channel realizations. Adding the rate bounds we obtain
\begin{eqnarray}
n\sum_{k=1}^KR_k&\leq&\frac{n}{2}\log(P)+\sum_{k=2}^K\left(H(\bar{Y}_{k-1}^{[n]}|G^n,W_{k},\cdots, W_K)-H(\bar{Y}_k^{[n]}|G^n,W_k, \cdots, W_K)\right)\nonumber\\
&&+n~o(\log(P))+o(n)
\end{eqnarray}
\begin{eqnarray}
&\leq&\frac{n}{2}\log(P)++n~o(\log(P))+o(n)\nonumber\\
&&+\sum_{k=2}^K\left[\max_{{w}_i\in\{W_i\},i\in[k:K]}\left(H(\bar{Y}_{k-1}^{[n]}|G^n,W_i={w}_i, \forall i\in[k:K])-H(\bar{Y}_k^{[n]}|G^n,W_i={w}_i, \forall i\in[k:K])\right)\right]\nonumber\\
\end{eqnarray}
In DoF terms,
\begin{eqnarray}
\bar{\mathcal{D}}_\Sigma &\leq&1 + \sum_{k\in[2:K]}\bar{\mathcal{D}}_{\Delta,k}\label{eq:final1}
\end{eqnarray}
So we need to bound each of the following difference of entropy terms, $\forall k\in[2:K]$
\begin{eqnarray}
\bar{\mathcal{D}}_{\Delta,k}&\triangleq&\limsup_{P\rightarrow\infty}\limsup_{n\rightarrow\infty}\max_{\substack{\mathbb{P}(\bar{X}_1^{[n]},\cdots, \bar{X}_K^{[n]})\\  (\bar{X}_1^{[n]},\cdots,\bar{X}_K^{[n]})\in\bar{\mathcal{X}}^{[n]}}}\frac{H(\bar{Y}_{k-1}^{[n]}|G^{[n]})-H( \bar{Y}_k^{[n]}|G^{[n]})}{\frac{n}{2}\log(P)} \label{eq:detkey}
\end{eqnarray} 
We will bound these terms one at a time. The remainder of the proof will show that $\bar{\mathcal{D}}_{\Delta,k}\leq \alpha_k$.
\item{\bf Functional Dependence $\bar{Y}_k^{[n]}(\bar{Y}_{k-1}^{[n]}, G^{[n]})$}\\
For a given channel realization for user $k-1$, $G_{k-1}^{[n]}$, there are multiple vectors $(\bar{X}_1^{[n]}, \bar{X}_2^{[n]},\cdots,\bar{X}_{k}^{[n]})$ that cast the same image in $\bar{Y}_{k-1}^{[n]}$. Thus, given the channel for user $k-1$, the mapping from $\bar{Y}_{k-1}^{[n]}$ to one of these vectors $(\bar{X}_1^{[n]}, \bar{X}_2^{[n]},\cdots,\bar{X}_{k}^{[n]})$ is random. Let us denote it by $\mathcal{L}$, i.e.,
\begin{eqnarray}
(\bar{X}_1^{[n]}, \bar{X}_2^{[n]},\cdots,\bar{X}_{k}^{[n]})&=&\mathcal{L}(\bar{Y}_{k-1}^{[n]}, G_{k-1}^{[n]})
\end{eqnarray}
Now note that
\begin{eqnarray}
H(\bar{Y}_k^{[n]}|G^n)&\geq&H(\bar{Y}_k^{[n]}|G^n,\mathcal{L})\\
&\geq&\min_{{L}\in\{\mathcal{L}\}}H(\bar{Y}_k^{[n]}|G^n,\mathcal{L}={L})
\end{eqnarray}
Let a minimizing mapping be ${L}_o$. Fix this as the deterministic mapping, 
\begin{eqnarray}
(\bar{X}_1^{[n]}, \bar{X}_2^{[n]},\cdots,\bar{X}_{k}^{[n]})&=&L_o(\bar{Y}_{k-1}^{[n]}, G_{k-1}^{[n]})
\end{eqnarray}
This implicitly allows the transmitter to have full knowledge of the channel vector of user $k-1$. We note that the choice of mapping does not affect the positive entropy term $H(\bar{Y}_{k-1}^{[n]}|G^{[n]})$ but it minimizes $H(\bar{Y}_k^{[n]}|G^{[n]})$, so that we can bound $\bar{\mathcal{D}}_{\Delta,k}$ as follows.

\begin{eqnarray}
\bar{\mathcal{D}}_{\Delta,k}&\leq&\hat{\mathcal{D}}_{\Delta,k}\triangleq\limsup_{P\rightarrow\infty}\limsup_{n\rightarrow\infty}\max_{\substack{\mathbb{P}(\bar{Y}_{k-1}^{[n]}|G^{[n]}), \bar{Y}_k^{[n]}(\bar{Y}_{k-1}^{[n]}, G^{[n]})\\  (\bar{X}_1^{[n]},\cdots,\bar{X}_K^{[n]})\in\bar{\mathcal{X}}^{[n]}}}\frac{H(\bar{Y}_{k-1}^{[n]}|G^{[n]})-H(\bar{Y}_{k}^{[n]}|G^{[n]})}{\frac{n}{2}\log(P)}\nonumber\\
&&\label{eq:final2}
\end{eqnarray}
Henceforth, note that $\bar{Y}_{k}^{[n]}$ is a function of $\bar{Y}_{k-1}^{[n]}, G^{[n]}$. 

\item {\bf Define Aligned Image Sets}\\
For channel realization $G^{[n]}$, define the aligned image set $S_{\bar{Y}_{k-1}^{[n]}}(G^{[n]})$ as the set of all $\bar{Y}_{k-1}^{[n]}$ that have the same image in $\bar{Y}_{k}^{[n]}$, i.e., 
\begin{eqnarray}
S_{\bar{\nu}^{[n]}}(G^{[n]})&\triangleq&\{\bar{y}_{k-1}^{[n]}\in\{\bar{Y}_{k-1}^{[n]}\}: \bar{Y}_k^{[n]}(\nu^{[n]}, G^{[n]})=\bar{Y}_k^{[n]}(\bar{y}_{k-1}^{[n]}, G^{[n]})\}
\end{eqnarray}

\item{\bf Bounding Difference of Entropies in Terms of Size of Aligned Image Sets}
\begin{eqnarray}
H(\bar{Y}_{k-1}^{[n]}|G^{[n]})&=&H(\bar{Y}_{k-1}^{[n]}, \bar{Y}_{k}^{[n]}|G^{[n]})\\
&=&H(\bar{Y}_{k}^{[n]}|G^{[n]})+H(\bar{Y}_{k-1}^{[n]}|{\bar{Y}_{k}^{[n]}},G^{[n]})\\
&=&H(\bar{Y}_{k}^{[n]}|G^{[n]})+H(S_{\bar{Y}_{k-1}^{[n]}}(G^{[n]})|G^{[n]})\\
&\leq&H(\bar{Y}_{k}^{[n]}|G^{[n]})+\mbox{E}\left[\log\left|S_{\bar{Y}_{k-1}^{[n]}}(G^{[n]})\right|\right]\\
&\leq&H(\bar{Y}_{k}^{[n]}|G^{[n]})+\log\mbox{E}\left[\left|S_{\bar{Y}_{k-1}^{[n]}}(G^{[n]})\right|\right]\label{eq:jensensk}
\end{eqnarray}
where we used Jensen's inequality in (\ref{eq:jensensk}). Rearranging terms, we note that
\begin{eqnarray}
\hat{\mathcal{D}}_{\Delta,k}&\leq&\limsup_{P\rightarrow\infty}\limsup_{n\rightarrow\infty}\max_{\substack{\mathbb{P}(\bar{Y}_{k-1}^{[n]}, G^{[n]}), \bar{Y}_k^{[n]}(\bar{Y}_{k-1}^{[n]},G^{[n]})\\  (\bar{X}_1^{[n]},\cdots,\bar{X}_K^{[n]})\in\bar{\mathcal{X}}^{[n]}}}\frac{\log\left(\mbox{E}\left[\left|S_{\bar{Y}_{k-1}^{[n]}}(G^{[n]})\right|\right]\right)}{\frac{n}{2}\log(P)}\label{eq:DSK}
\end{eqnarray}
\item{\bf Bounding the Probability of Image Alignment}\\
Given the channel, $G^{[n]}_{k-1}$, of user $k-1$ and two realizations of $\bar{Y}_{k-1}^{[n]}$, say $\bar{y}^{[n]}$ and $\bar{y'}^{[n]}$, which map to $\bar{X}_j^{[n]}(\bar{y}^{[n]}, G_{k-1}^{[n]})=\bar{x}_j^{[n]}$ and $\bar{X}_j^{[n]}(\bar{y'}^{[n]},G_{k-1}^{[n]})=\bar{x'}_j^{[n]}$, $\forall j\in[1:k]$, let us bound the probability that they produce the same image $\bar{Y}_k^{[n]}$. For notational compactness let us define $G_{kk}(t)=1, \forall k\in[1:K], \forall t\in[1:n]$. Note that for $\bar{y'}^{[n]}\in S_{\bar{y}^{[n]}}(G^{[n]})$ we must have, $\forall t\in[1:n]$
\begin{eqnarray}
\sum_{j\in[1:k]}\lfloor G_{kj}(t)\bar{x'}_j(t)\rfloor &=&\sum_{j\in[1:k]}\lfloor G_{kj}(t)\bar{x}_j(t)\rfloor \\
\Rightarrow \lfloor G_{kj^*(t)}(t)\bar{x'}_{j^*(t)}(t)\rfloor-\lfloor G_{kj^*(t)}(t)\bar{x}_{j^*(t)}(t)\rfloor &=&\sum_{\substack{j\in[k:k],j\neq j^*(t)}}\left(\lfloor G_{kj}(t)\bar{x}_j(t)\rfloor -\lfloor G_{kj}(t)\bar{x'}_j(t)\rfloor\right)\\
\Rightarrow  G_{kj^*(t)}(t)\left(\bar{x'}_{j^*(t)}(t)-\bar{x}_{j^*(t)}(t)\right) &\in&\sum_{j\in[1:k],j\neq j^*(t)}\left(\lfloor G_{kj}(t)\bar{x}_j(t)\rfloor -\lfloor G_{kj}(t)\bar{x'}_j(t)\rfloor\right)+\Delta_{(-1,1)}\nonumber\\
\end{eqnarray}
where $\Delta_{(-1,1)}\in(-1,1)$, and we define
\begin{eqnarray}
j^*(t)&\triangleq& \arg \max_{j\in[1:k-1]}|\bar{x'}_{j}(t)-\bar{x}_{j}(t)|
\end{eqnarray}
Thus, for all $t\in[1:n]$ such that $\bar{x'}_{j^*(t)}(t)\neq\bar{x}_{j^*(t)}(t)$, the value of $G_{kj^*(t)}(t)$ must lie within an interval of length no more than $\frac{2}{\left| \bar{x'}_{j^*}(t)-\bar{x}_{j^*}(t)\right|}$. Therefore,  the probability that the images due to $\bar{y}^{[n]}$ and $\bar{y'}^{[n]}$ align at user $k$, is bounded as follows.
\begin{eqnarray}
\mathbb{P}\left(\bar{y'}^{[n]}\in S_{\bar{y}^{[n]}}(G^{[n]})\right)&\leq&f_{\max,k}^n\prod_{t:\bar{x'}_{j^*(t)}(t)\neq\bar{x}_{j^*(t)}(t)}\frac{2}{\left| \bar{x'}_{j^*(t)}(t)-\bar{x}_{j^*(t)}(t)\right|}
\end{eqnarray}
It will be useful to express the bound in terms of $\bar{y'}(t), \bar{y}(t)$. To this end, let us proceed as follows. 
\begin{eqnarray}
\bar{y'}(t)-\bar{y}(t)&=&\sum_{j=1}^{k-1}\left(\lfloor G_{k-1,j}(t)\bar{x'}_j(t)\rfloor-\lfloor G_{k-1,j}(t)\bar{x}_j(t)\rfloor\right)\\
&\in&\sum_{j=1}^{k-1}\left(\lfloor G_{k-1,j}(t)\left(\bar{x'}_j(t)- \bar{x}_j(t)\right)\rfloor\right) + (-K,K)\\
|\bar{y'}(t)-\bar{y}(t)|&\leq&|\bar{x'}_{j^*(t)}(t)- \bar{x}_{j^*(t)}(t)|\sum_{j=1}^{k-1}| G_{k-1,j}(t)| + K\\
\Rightarrow \frac{1}{|\bar{x'}_{j^*(t)}(t)- \bar{x}_{j^*(t)}(t)|}&\leq&\frac{\sum_{j=1}^{k-1}| G_{k-1,j}(t)|}{|\bar{y'}(t)-\bar{y}(t)|-K}
\end{eqnarray}
whenever $|\bar{y'}(t)-\bar{y}(t)|>K$. Therefore,
\begin{eqnarray}
\mathbb{P}\left(\bar{y'}^{[n]}\in S_{\bar{y}^{[n]}}(G^{[n]})\right)&\leq&\bar{g}^n(f_{\max,k})^n\prod_{t:|\bar{y'}(t)-\bar{y}(t)|>K}\frac{1}{|\bar{y'}(t)-\bar{y}(t)|-K}
\end{eqnarray}
where 
\begin{eqnarray}
\bar{g}^n&\triangleq&\max\left(1,\prod_{t:\bar{x'}_{j^*(t)}(t)\neq\bar{x}_{j^*(t)}(t)}2\sum_{j=1}^{k-1}| G_{k-1,j}(t)| \right)
\end{eqnarray}
\item{\bf Bounding the Average Size of Aligned Image Sets}
\begin{eqnarray}
\mbox{E}\left[S_{\bar{y}}^{[n]}(G^n)\right]&=&\sum_{\bar{y'}^{[n]}\in\{{\bar{Y}_{k-1}^{[n]}}\}}\mathbb{P}\left(\bar{y'}^{[n]}\in S_{\bar{y}^{[n]}}(G^n)\right)\\
&\leq&\bar{g}^n(f_{\max,k})^n\prod_{t=1}^n\left(\sum_{\bar{y'}(t): |\bar{y'}(t)-\bar{y}(t)|\leq K}1+\sum_{\bar{y'}(t): K< |\bar{y'}(t)-\bar{y}(t)|\leq Q_y(t)}\frac{1}{|\bar{y'}(t)-\bar{y}(t)|-K}\right)\nonumber\\
&&\\
&\leq&\bar{g}^n(f_{\max,k})^n\left(\log(\sqrt{P})+o(\log(P))\right)^n\label{eq:aveSK}
\end{eqnarray}
where $Q_y(t)\leq\lceil\sqrt{P}\rceil\sum_{j\in[1:k-1]}(|G_{k-1,j}(t)|+K)$. 

\item{\bf Combining the Bounds to Complete the Proof}\\
Combining (\ref{eq:aveSK}) and (\ref{eq:DSK}) we have
\begin{eqnarray}
\hat{\mathcal{D}}_{\Delta,k}&\leq&\limsup_{P\rightarrow\infty}\limsup_{n\rightarrow\infty}\frac{\log\left((\bar{g}f_{\max,k})^n(\log(\sqrt{P})+o(\log(P)))^n\right)}{\frac{n}{2}\log(P)}\\
&\leq&\alpha_{k}\label{eq:finalfinal}
\end{eqnarray}
Finally, combining (\ref{eq:final1}), (\ref{eq:final2}) and (\ref{eq:finalfinal}) we have the result,
\begin{eqnarray}
\bar{\mathcal{D}}_\Sigma &\leq &1+\alpha_2+\cdots+\alpha_K
\end{eqnarray}
\end{enumerate}

\section{Discussion}
Since CSIT is almost never available with infinite precision, the collapse of DoF under finite precision channel uncertainty is a sobering result that stands in stark contrast against the  tremendous DoF gains shown to be possible with perfect channel knowledge \cite{Cadambe_Jafar_int, Cadambe_Jafar_X}. However, as evident from the conjecture of Lapidoth, Shamai and Wigger,  the pessimistic outcome is not unexpected. In terms of practical implications, just like the extremely positive DoF results, the extremely negative DoF results should be taken with a grain of salt. The collapse of DoF under finite precision CSIT is very much due to the asymptotic nature of the DoF metric, and may not be directly representative of finite SNR scenarios which are of primary concern in practice. From a technical perspective, the new outer bound technique offers hope for new insights through the studies of more general forms of CSIT, such as finite precision versions of delayed \cite{Maddah_Tse}, mixed \cite{Sheng_Kobayashi_Gesbert_Yi,Gou_Jafar}, topological \cite{Jafar_TIM}, blind \cite{Jafar_corr} and alternating \cite{Tandon_Jafar_Shamai_Poor} CSIT.

\bigskip

\appendix
\noindent{\bf \LARGE Appendix}
\section{Proof of Lemma \ref{lemma:detn}}\label{app:detn}
The proof is in two parts. First we prove that we can limit the inputs and outputs to integer values without reducing the DoF. Then, we will show that the long-term (per-codeword) power constraints can be replaced with short-term (per-symbol) power constraints without reducing the DoF. The proofs follow along the lines of similar proofs by Bresler and Tse in \cite{Bresler_Tse}, are specialized to the broadcast setting, and fill in several details that are omitted in\cite{Bresler_Tse}. 
\subsection{Integer Inputs and Outputs}
Given codebooks with real codewords, $(X_1^{[n]}, X_2^{[n]})\in\mathbb{R}^{n}\times\mathbb{R}^{n}$ for the canonical channel model, we show that the deterministic channel model with integer inputs $\lfloor X_1^{[n]}\rfloor, \lfloor X_2^{[n]}\rfloor$,  and outputs 
\begin{eqnarray}
\bar{\bar{Y}}_1(t)&\triangleq & \lfloor X_1(t)\rfloor\\
\bar{\bar{Y}}_2(t)&\triangleq & \lfloor G(t)\lfloor X_1(t)\rfloor \rfloor+\lfloor X_2(t)\rfloor
\end{eqnarray}
achieves  the same DoF. Thus, removing noise and limiting the inputs to integer values, as done in the deterministic model, does not reduce DoF relative to the original canonical channel model.

\noindent Define $E_1^{[n]}={X}_{1}^{[n]}-\left \lfloor {X}_{1}^{[n]} \right \rfloor+Z_1^{[n]}$. Taking a similar approach to Lemma 5 in \cite{Bresler_Tse} we have,
\begin {eqnarray}
I(W_1;Y_1^{[n]}|G^{[n]})&=&I(W_1;\bar{\bar{Y}}_1^{[n]}+E_1^{[n]}|G^{[n]})\\
&\leq&I(W_1;\bar{\bar{Y}}_1^{[n]}, E_1^{[n]}|G^{[n]})\\
&\leq&I(W_1;\bar{\bar{Y}}_1^{[n]}|G^{[n]})+I(W_1;E_1^{[n]}|G^{[n]}, \bar{\bar{Y}}_1^{[n]})\\
&\leq&I(W_1;\bar{\bar{Y}}_1^{[n]}|G^{[n]})+h(E_1^{[n]}|G^{[n]})-h(E_1^{[n]}|G^{[n]},\bar{\bar{Y}}_1^{[n]},X_1^{[n]})\\
&\leq&I(W_1;\bar{\bar{Y}}_1^{[n]}|G^{[n]})+h(E_1^{[n]}|G^{[n]})-h(Z_1^{[n]})\\
&\leq&I(W_1;\bar{\bar{Y}}_1^{[n]}|G^{[n]})+ \frac{n}{2}\log(2)
\end{eqnarray}
so that the difference between $I(W_1;Y_1^{[n]}|G^{[n]})$ and $I(W_1;\bar{\bar{Y}}_1^{[n]}|G^{[n]})$ approaches $0$ when normalized by $\frac{n}{2}\log(P)$, as first $n$ and then $P$ is sent to infinity.

Similarly, by defining $E_2^{[n]}=G^{n} X_1^{[n]} -\left \lfloor G^{n} \left \lfloor X_1^{[n]} \right \rfloor \right \rfloor+X_2^{[n]}-\left \lfloor X_2^{[n]} \right \rfloor+Z_2^{[n]}$, we have,
\begin {eqnarray}
I(W_2;Y_2^{[n]}|G^{[n]})&=&I(W_2;\bar{\bar{Y}}_2^{[n]}+E_2^{[n]}|G^{[n]})\\
&\leq&I(W_2;\bar{\bar{Y}}_2^{[n]}|G^{[n]})+h(E_2^{[n]}|G^{[n]})-h(Z_2^{[n]})\\
& \leq& I(W_2;\bar{\bar{Y}}_2^{[n]}|G^{[n]})+ \sum_{t=1}^n \mbox{E}_{G(t)}\left[\frac{1}{2}\log\left((G(t)+1)^2+1 \right) \right]
\end{eqnarray}
so that the difference between $I(W_2;Y_2^{[n]}|G^{[n]})$ and $I(W_2;\bar{\bar{Y}}_2^{[n]}|G^{[n]})$ approaches $0$ when normalized by $\frac{n}{2}\log(P)$, as first $n$ and then $P$ is sent to infinity.. Thus, the deterministic channel   with inputs $(\lfloor X_1^{[n]}\rfloor, \lfloor X_2^{[n]}\rfloor)$, outputs $(\bar{\bar{Y}}_1^{[n]}, \bar{\bar{Y}}_2^{[n]})$, and per-codeword power constraints, 
\begin{eqnarray}
\frac{1}{n}\sum_{t=1}^n\left((\lfloor{X}_1(t)\rfloor)^2+(\lfloor{X}_2(t)\rfloor)^2\right)&\leq& P
\end{eqnarray}
achieves at least the same DoF as the original canonical channel model.

\subsection{Per-symbol Power Constraints}
Given codewords $(\lfloor X_1^{[n]}\rfloor, \lfloor X_2^{[n]}\rfloor)$ for the deterministic channel with outputs $(\bar{\bar{Y}}_1^{[n]}, \bar{\bar{Y}}_2^{[n]})$, such that the codewords satisfy per-codeword power constraints, define $\forall t\in[1:n]$
\begin{eqnarray}
\bar{X}_1(t)&=&\lfloor X_1(t)\rfloor \mbox{ mod } \lceil \sqrt{P} \rceil\\
\bar{X}_2(t)&=&\lfloor X_2(t)\rfloor \mbox{ mod } \lceil \sqrt{P} \rceil\\
\bar{Y}_1(t)&=&\bar{X}_1(t)\\
\bar{Y}_2(t)&=&\lfloor G(t)\bar{X}_1(t)\rfloor+\bar{X}_2(t)\\
\hat{X}_1(t)&=&\lfloor X_1(t)\rfloor -\bar{X}_1(t)\\
\hat{X}_2(t)&=&\lfloor X_2(t)\rfloor -\bar{X}_2(t)\\
\hat{Y}_1(t)&=&\bar{\bar{Y}}_1(t) -\bar{Y}_1(t)\\
\hat{Y}_2(t)&=&\bar{\bar{Y}}_2(t)- \bar{Y}_2(t)
\end{eqnarray}
Since $\bar{X}_1(t), \bar{X}_2(t)\in\{0,1,\cdots, \lceil\sqrt{P}\rceil\}$,  the new codewords $(\bar{X}_1^{[n]}, \bar{X}_2^{[n]})$ satisfy per-symbol power constraints. Now let us compare the rates achieved on the channel $(\lfloor X_1^{[n]}\rfloor, \lfloor X_2^{[n]}\rfloor)\longrightarrow (\bar{\bar{Y}}_1^{[n]}, \bar{\bar{Y}}_2^{[n]})$ to the rates achieved on the new channel $(\bar{X}_1^{[n]}, \bar{X}_2^{[n]})\longrightarrow (\bar{Y}_1^{[n]}, \bar{Y}_2^{[n]})$.
\begin{eqnarray}
I(W_1; \bar{\bar{Y}}_1^{[n]}|G^{[n]})&=&I(W_1;\bar{Y}_1^{[n]}+\hat{Y}_1^{[n]}|G^{[n]})\\
&\leq&I(W_1;\bar{Y}_1^{[n]},\hat{Y}_1^{[n]}|G^{[n]})\\
&\leq&I(W_1;\bar{Y}_1^{[n]}|G^{[n]})+H(\hat{Y}_1^{[n]}|G^{[n]})\\
&\leq& I(W_1;\bar{Y}_1^{[n]}|G^{[n]})+\sum_{t=1}^n H(\hat{Y}_1(t))\\
&\leq& I(W_1;\bar{Y}_1^{[n]}|G^{[n]})+\sum_{t=1}^n H(\hat{X}_1(t))
\end{eqnarray}
Similarly
\begin{eqnarray}
I(W_2; \bar{\bar{Y}}_2^{[n]}|G^{[n]})&=&I(W_2;\bar{Y}_2^{[n]}+\hat{Y}_2^{[n]}|G^{[n]})\\
&\leq&I(W_2;\bar{Y}_2^{[n]},\hat{Y}_2^{[n]}|G^{[n]})\\
&\leq&I(W_2;\bar{Y}_2^{[n]}|G^{[n]})+H(\hat{Y}_2^{[n]}|G^{[n]})\\
&\leq& I(W_2;\bar{Y}_2^{[n]}|G^{[n]})+\sum_{t=1}^n H(\hat{Y}_2(t)|G(t))\\
&=&I(W_2;\bar{Y}_2^{[n]}|G^{[n]})+\sum_{t=1}^n H( \lfloor G(t)\lfloor X_1(t)\rfloor \rfloor+\lfloor X_2(t)\rfloor - \lfloor G(t)\bar{X}_1(t)\rfloor-\bar{X}_2(t)|G(t))\nonumber\\
&&\\
&=&I(W_2;\bar{Y}_2^{[n]}|G^{[n]})+\sum_{t=1}^n H( \lfloor G(t)\hat{X}_1(t) \rfloor+\hat{X}_2(t) +\Delta(t)|G(t))\\
&\leq&I(W_2;\bar{Y}_2^{[n]}|G^{[n]})+\sum_{t=1}^n H(\lfloor G(t)\hat{X}_1(t) \rfloor|G(t))+\sum_{t=1}^nH(\hat{X}_2(t))+\sum_{t=1}^nH(\Delta(t))\\
&\leq&I(W_2;\bar{Y}_2^{[n]}|G^{[n]})+\sum_{t=1}^n H(\hat{X}_1(t))+\sum_{t=1}^n H(\hat{X}_2(t))+n\log(3)
\end{eqnarray}
where $\Delta(t)\in\{-1,0,1\}$. To complete the proof we only need to show that $\sum_{t=1}^n H(\hat{X}_i(t))\leq n~o(\log(P))$, i.e., these terms can contribute no more than 0 in the DoF sense. We will use the fact that any integer number $X$ can be written as $Q\lfloor\frac{X}{Q}\rfloor - Q1(x<0) + (X \mod Q)$ where $Q\triangleq\lceil \sqrt{P} \rceil$ is also an integer value.
\begin{eqnarray}
H(\hat{X}_1(t))&=&H\left(Q\left\lfloor\frac{1}{Q}\lfloor{X}_1(t)\rfloor\right\rfloor - Q\mathbb{I}(\lfloor X_1(t)\rfloor <0)\right)\\
&=&H\left(\left\lfloor\frac{1}{Q}\lfloor{X}_1(t)\rfloor\right\rfloor - \mathbb{I}(\lfloor X_1(t)\rfloor <0)\right)\\
&\leq&H\left(\left\lfloor\frac{1}{Q}\lfloor{X}_1(t)\rfloor\right\rfloor\right) + H\left( \mathbb{I}(\lfloor X_1(t)\rfloor <0)\right)\\
&\leq&H\left(\left\lfloor\frac{1}{Q}\lfloor{X}_1(t)\rfloor\right\rfloor\right) +1\label{eq:X1byQ}
\end{eqnarray}
Define
\begin{eqnarray}
p(t) &=&\frac{\mbox{E}\left(\lfloor X_1(t)\rfloor\right)^2}{nP}
\end{eqnarray}
where the expectation is over messages, i.e., the choice of codewords, so that
\begin{eqnarray}
\sum_{t=1}^n p(t)&\leq&1\label{eq:sumpt}
\end{eqnarray}
Let $\rho_k\triangleq \mathbb{P}(\lfloor X_1(t)\rfloor = k)$, so that
\begin{eqnarray}
\sum_{k=-\infty}^\infty k^2\rho_k&\leq&p(t)nP
\end{eqnarray} 
Next let us bound the probability $\tilde\rho_m$ of $\lfloor X_1(t)\rfloor$ falling within the $m^{th}$ quantization interval.
\begin{eqnarray}
\tilde\rho_m&\triangleq& \sum_{k=mQ}^{(m+1)Q-1}\rho_k, ~~\forall m\in\mathbb{Z}\\
\tilde\rho_m&\leq&\frac{1}{\min(m^2,(m+1)^2)Q^2}\sum_{k=mQ}^{(m+1)Q-1}k^2\rho_k, ~\forall m\in\mathbb{Z}/ \{0,-1\}\\
&\leq& \frac{np(t)}{\min(m^2,(m+1)^2)}, ~\forall m\in\mathbb{Z}/ \{0,-1\}\\
{\tilde{\rho}_m^*}&\triangleq&\frac{np(t)}{\min(m^2,(m+1)^2)}
\end{eqnarray}

In the following derivation we will make use of the fact that $\tilde{\rho}_m\log\left(\frac{1}{\tilde{\rho}_m}\right)$ is an increasing function of $\tilde{\rho}_m$ when $\tilde{\rho}_m<1/e$, so for these $\tilde{\rho}_m$ values one can replace $\tilde{\rho}_m$ with $\tilde{\rho}_m^*$ to obtain an outer bound. Further, we will use the fact that the maximum value of $\tilde{\rho}_m\log\left(\frac{1}{\tilde{\rho}_m}\right)$ is $\frac{1}{e\ln(2)}$. 

We bound the entropy term in (\ref{eq:X1byQ}) as follows
\begin{eqnarray}
H\left(\left\lfloor\frac{1}{Q}\lfloor{X}_1(t)\rfloor\right\rfloor\right) &=&\sum_{m=-\infty}^{\infty}\tilde{\rho}_m\log\left(\frac{1}{\tilde{\rho}_m}\right)\\
&=&\sum_{m\in\{0,-1,1\}}\tilde{\rho}_m\log\left(\frac{1}{\tilde{\rho}_m}\right)+\sum_{\substack{m: \tilde{\rho}_m^*> \frac{1}{e}\\m\notin\{0,-1,1\}}}\tilde{\rho}_m\log\left(\frac{1}{\tilde{\rho}_m}\right)+\sum_{\substack{m: \tilde{\rho}_m^*\leq \frac{1}{e}\\m\notin\{0,-1,1\}}}\tilde{\rho}_m\log\left(\frac{1}{\tilde{\rho}_m}\right)\nonumber\\
&&\\
&=&\sum_{m\in\{0,-1,1\}}\frac{1}{e\ln(2)}+\sum_{\substack{m: \tilde{\rho}_m^*> \frac{1}{e}\\m\notin\{0,-1,1\}}}\frac{1}{e\ln(2)}+\sum_{\substack{m: \tilde{\rho}_m^*\leq \frac{1}{e}\\m\notin\{0,-1,1\}}}\tilde{\rho}_m^*\log\left(\frac{1}{\tilde{\rho}_m^*}\right)\\
&\leq& \frac{3}{e\ln(2)}+\frac{4\sqrt{np(t)}}{e\ln(2)}+2\sum_{m=2}^\infty \frac{np(t)}{m^2}\log\left(\frac{m^2}{np(t)}\right)\\
&\leq& \frac{3+4\sqrt{np(t)}}{e\ln(2)}+4np(t)\sum_{m=2}^\infty \frac{\log(m)}{m^2}+2np(t)\log\left(\frac{1}{np(t)}\right)\sum_{m=2}^\infty \frac{1}{m^2}\label{eq:summations}\\
&\leq&\frac{3+4\sqrt{np(t)}}{e\ln(2)}+6np(t)+2np(t)\log\left(\frac{1}{np(t)}\right)\\
&\leq&\frac{3+4\max(1, np(t))}{e\ln(2)}+6np(t)+\frac{2}{e\ln(2)}\label{eq:xlogx}
\end{eqnarray}
where in (\ref{eq:summations}) we used the facts that $\sum_{m=2}^\infty \frac{\log(m)}{m^2}\leq 1.5$ and $\sum_{m=2}^\infty\frac{1}{m^2}\leq 1$. In (\ref{eq:xlogx}) we used the property that the maximum value of the function $x\log\left(\frac{1}{x}\right)$ is $\frac{1}{e\ln(2)}$.

Substituting into  (\ref{eq:X1byQ}), summing over all $t\in[1:n]$, and using (\ref{eq:sumpt}) we have
\begin{eqnarray}
\sum_{t=1}^nH(\hat{X}_1(t))&\leq& n + \frac{5n}{e\ln(2)}+ \frac{4}{e\ln(2)}\sum_{t=1}^n(1+np(t))+6n\sum_{t=1}^np(t)\\
&\leq&n \left(1+\frac{13}{e\ln(2)}+6\right)\\
&\leq& n ~o(\log(P))
\end{eqnarray}
The same arguments show that $\sum_{t=1}^nH(\hat{X}_2(t))\leq n ~o(\log(P))$ as well. Thus, we conclude that the deterministic channel with per-symbol power constraints, achieves at least the same DoF as the deterministic channel with per-codeword power constraints.

\section{Channel Model: Reduction to Canonical Form}\label{app:z}
Define
\begin{eqnarray}
X_1(t)&=&\tilde{G}_{11}(t)\tilde{X}_1(t)+{\tilde{G}_{12}(t)}\tilde{X}_2(t)\\
X_2(t)&=&\left(\frac{\mbox{det}({\bf \tilde{G}}(t))}{\tilde{G}_{11}(t)}\right)\tilde{X}_2(t)\\
G(t)&=&\frac{\tilde{G}_{21}(t)}{\tilde{G}_{11}(t)}\\
P&=&(2{M}^2+{M}^4)\tilde{P}
\end{eqnarray}
Note that $P=\Theta(\tilde{P})$, and $G(t)$ is  bounded away from zero and infinity, as $\frac{1}{{M}^2}<|G(t)|<{M}^2$, $\forall t\in\mathbb{N}$. 
\begin{eqnarray}
(X_1(t))^2+(X_2(t))^2&\leq&\left((\tilde{G}_{11}(t))^2+(\tilde{G}_{12}(t))^2+\left(\frac{\mbox{det}({\bf \tilde{G}}(t))}{\tilde{G}_{11}(t)}\right)^2\right)((\tilde{X}_1(t))^2+(\tilde{X}_2(t))^2)\\
&\leq& (2{M}^2+{M}^4)((\tilde{X}_1(t))^2+(\tilde{X}_2(t))^2)
\end{eqnarray}

In addition to the channel vector for user 1, let us allow the CSIT to include the determinant of the channel matrix. This cannot reduce capacity, so it can only make the outer bound stronger.
\begin{eqnarray}
\tilde{G}_{11}(t), \tilde{G}_{12}(t), \mbox{det}({\bf G}(t))\in\mathcal{T}, &&\forall t\in\mathbb{N}
\end{eqnarray}
Note that for continuous distributions, $G(t)$ is not a function of $\mathcal{T}$. 
With the available CSIT, suppose the transmitter sets:
\begin{eqnarray}
\tilde{X}_1(t)&=&\frac{1}{\tilde{G}_{11}(t)}X_1(t)-\left(\frac{\tilde{G}_{12}(t)}{\mbox{det}({\bf \tilde{G}}(t))}\right)X_2(t)\\
\tilde{X}_2(t)&=&\left(\frac{\tilde{G}_{11}(t)}{\mbox{det}({\bf \tilde{G}}(t))}\right)X_2(t)
\end{eqnarray}
Substituting into (\ref{eq:channelmodel}) we obtain the canonical channel model (\ref{eq:canonical}). Noting that the transformation from $\tilde{X}_1(t), \tilde{X}_2(t)$ to $X_1(t), X_2(t)$ is invertible and that the new power constraint (\ref{eq:powercanonical}) allows all feasible $\tilde{X}_1(t), \tilde{X}_2(t)$, it is evident that the capacity of the channel in its canonical form cannot be smaller than that of the original channel. Thus, the canonical channel transformation is valid for our DoF outer bound.
\hfill$\Box$

{\it Remark:} It is not necessary to provide side information of the determinant of the channel matrix to the transmitter. One could also normalize the desired channel coefficient values to unity by scaling the received signals at the receivers, which would only scale the noise variance by a bounded amount that is inconsequential for DoF. We choose to provide the determinant as side information to the transmitter, because for a pessimistic outer bound that shows the collapse of DoF, including more CSIT only makes the result stronger. It shows that even this additional CSIT cannot prevent the collapse of DoF.

\section*{Acknowledgments}
S.  Jafar would like to gratefully acknowledge many insightful discussions over the years with Prof. Shlomo Shamai regarding the problem studied in this work.
\bibliographystyle{IEEEtran}
\bibliography{Thesis}
\end{document}